\documentclass[12pt,preprint]{emulateapj}

\begin{document}

\title{A continuum of accretion burst behavior in young stars observed by $K2$}

\author{
Ann Marie Cody\altaffilmark{1},
Lynne A. Hillenbrand\altaffilmark{2},
Trevor J. David\altaffilmark{2},
John M. Carpenter\altaffilmark{2,3},
Mark E. Everett\altaffilmark{4},
Steve B. Howell\altaffilmark{1}
}

\altaffiltext{1}{NASA Ames Research Center, Moffet Field, CA 94035}
\altaffiltext{2}{Department of Astronomy, California Institute of Technology, Pasadena CA 91125}
\altaffiltext{3}{current address: Joint ALMA Observatory, Av. Alonso de C\'{o}rdova 3107, Vitacura, Santiago, Chile}
\altaffiltext{4}{National Optical Astronomy Observatory, 950 N. Cherry Ave., Tucson, AZ 85719, USA}

\begin{abstract}
We present 29 likely members of the young $\rho$ Oph or Upper Sco regions of recent star formation that exhibit ``accretion burst" type light curves in $K2$ time series photometry. The bursters were identified by visual examination of their $\sim$80 day light curves, though all satisfy the $M < -0.25$ flux asymmetry criterion for burst behavior defined by Cody et al. (2014). The burst sources represent $\approx$9\% of cluster members with strong infrared excess indicative of circumstellar material. Higher amplitude burster behavior is correlated with larger inner disk infrared excesses, as inferred from {\em WISE} $W1-W2$ color. The burst sources are also outliers in their large H$\alpha$ emission equivalent widths.  No distinction between bursters and non-bursters is seen in stellar properties such as multiplicity or spectral type. The frequency of bursters is similar between the younger, more compact $\rho$ Oph region, and the older, more dispersed Upper Sco region.  The bursts exhibit a range of shapes, amplitudes ($\sim$10-700\%), durations ($\sim$1-10 days), repeat time scales ($\sim$3-80 days), and duty cycles ($\sim$10-100\%).  Our results provide important input to models of magnetospheric accretion, in particular by elucidating the properties of accretion-related variability in the low state between major longer duration events such as EX Lup and FU Ori type accretion outbursts.  We demonstrate the broad continuum of accretion burst behavior in young stars -- extending the phenomenon to lower amplitudes and shorter timescales than traditionally considered in the theory of pre-main sequence accretion history.

\end{abstract}

\keywords{}
  
\section{Introduction}

Variable mass flux has long been recognized as an important element of protostellar and pre-main sequence accretion. Accretion rates are believed to be higher ($\sim 10^{-5} M_\odot$~yr$^{-1}$) during the first 10$^5$ years of protostellar evolution, with frequent outbursts of up to $10^{-4} M_\odot$~yr$^{-1}$ \citep{1993prpl.conf..497H}. 
The bursts are predicted as a consequence of unstable pile-up of gas in the inner disk, which then intermittently releases a cascade of material onto the star due to viscous-thermal disk instabilities \citep{1994ApJ...427..987B,2009ApJ...694.1045Z,2010MNRAS.406.1208D}.  The burst frequency and perhaps amplitude decline over time \citep[e.g.,][]{1996ARA&A..34..207H,2015ApJ...805..115V}.  

While most of the stellar mass is thought to accumulate in the early phases of protostellar evolution, accretion at rates of  $10^{-10}$--$10^{-6} M_\odot$~yr$^{-1}$ persists through the T Tauri phase (ages up to a few Myr), with less frequent bursts \citet{hhc2016}. The currently accepted picture of T Tauri star accretion involves magnetic funnel flows channeling gas from the inner disk onto the central star. Where this material impacts the surface, shocks arise and thermal hot spots form. There may be one spot near each magnetic pole, or multiple spot complexes that are distributed about the stellar surface. The number and geometry of the funnel flows is thought to depend on the accretion rate \citep[e.g.,][]{2008ApJ...673L.171R}.

Empirically, time series monitoring of young stellar objects (YSOs; ages $<$1--10~Myr) has an extensive history. It has been known since or before e.g. \cite{1949ApJ...110..424J} that T~Tauri stars display flux variations at a wide variety of timescales and magnitudes. The outbursting FU Ori stars and their lower amplitude, repeating cousins the EX Lup stars, are at the extreme end of the variabiity spectrum -- and rare. More common photometric variability is characterized by smaller amplitude and shorter timescale fluctuations. The studies of \citet{1994AJ....108.1906H}, \citet{2007A&A...461..183G},   \citet{2008MNRAS.391.1913R}, and \citet{2013ApJ...768...93F} have illustrated and quantified much of the ``typical" young star photometric phenomena occurring from sub-hour to multi-decade time scales. One cause of the routine brightness variations is sporadic infall of material from the surrounding disk.

Even in their predominant low-state accretion phases, young stars are understood as variable accretors.  The photometric variability is complemented by highly variable emission line profiles and veiling \citep[e.g.][]{2013AJ....145..108C,2014MNRAS.440.3444C} including in stars with relatively low accretion rates such as TW Hya and V2129 Oph \citep{2002ApJ...571..378A,2012A&A...541A.116A}. 

Nevertheless, the variability time scales and accretion rate changes remain poorly quantified for typical young accreting star/disk systems. There may be a continuum of ``burst" behavior with a range of amplitudes and time scales that have not yet been appropriately sampled or appreciated in existing ground-based data sets. While most of the stellar mass has been assumed to accumulate in the episodic and dramatic fashion of the rare large FU Ori and Ex Lup type events, the role and implications of discrete lower amplitude accretion events \citep[e.g.,][]{2014AJ....147...83S} and continuously stochastic accretion behavior \citep[e.g.,][]{2016AJ....151...60S} is not well understood in the context of stellar mass accumulation and inner disk evolution.  

In probing accretion variability, space-based photometric campaigns have several advantages over ground-based work, including near-continuous sampling (versus interruptions for daytime, weather, etc), higher measurement precision, and fainter signal detection limits. A detailed analysis of optical and infrared variability among disk-bearing stars in the $\sim$3~Myr NGC~2264 was conducted by \cite{2014AJ....147...82C} based on a forty-day optical time series from the {\em CoRoT} space telescope at 10-minute cadence, along with 30 days of {\em Spitzer} Space Telescope monitoring at 100-minute cadence. These data enabled an unprecedented view of YSO brightness changes on a variety of timescales. Among the detected variability groups was a new class of ``stochastic accretion burst'' light curves -- dominated by brightening events
of duration 0.1-1 days and amplitude 5-50\% the quiescent flux value \citep{2014AJ....147...83S}. It was speculated that these events were caused by the unsteady infall of material onto the stellar surface, as predicted by \citet{2011MNRAS.416..416R}. They may thus represent ``normal" discrete accretion variations and bursts, in contrast to the FU Ori and EX Lup outbursts described above.

The NASA $K2$ mission Campaign 2 observations included the young $\rho$~Ophiuchus molecular cloud region at $<$1-2~Myr, and the adjacent Upper Scorpius OB association, which is debated from analysis of HR diagrams to be either $\sim$3-5~Myr based on the low mass stellar population \citep[e.g.,][]{2002AJ....124..404P,2015ApJ...808...23H} or $\sim$11~Myr based on the solar and super-solar mass population \citep{2012ApJ...746..154P}; the latter age is beginning to be favored by results on eclipsing binaries \citep{2015ApJ...807....3K,2016ApJ...816...21D} and asteroseismology \citep{2015MNRAS.454.2606R}. By sampling stars with ages comparable to and extending to much older than NGC 2264 (in $\rho$~Oph and Upper Sco, respectively), the $K2$ time series data can be used to compare accretion burst behavior as a function of age and therefore presumably disk properties which are expected to evolve with time.

We report here on a continuum of accretion burst behavior among members of $\rho$~Oph and Upper Sco. 
We observe discrete brightening events that range in amplitude from 0.1 to 2.5 magnitudes 
and in timescale from $<$1 day to $>$1 week. We demonstrate that the burst phenomenon is seen only in those stars with evidence for strongly accreting disks, distinct from the typical disks in the region with weaker infrared excess and H$\alpha$ emission. Section 2 contains description of the $K2$ observations
and pixel file processing, and of follow-up high dispersion spectroscopy. Section 3 presents the light curve analysis and identification of burst type variables,
Section 4 a discussion of the corresponding accretion and disk properties, and Section 5 the spatial and time domain characteristics of bursting sources. We discuss the implications of these observations in Section~6 and summarize the results in Section~7.

\section{Observations}
\subsection{$K2$ Photometry}
  
The $K2$ mission \citep{2014PASP..126..398H} observed nearly 2000 stars in the young $\rho$~Ophiuchus and Upper Scorpius regions during Campaign 2. We have mainly considered objects submitted under programs GO2020, GO2047, GO2052, GO2056, GO2063, and GO2085 of the Campaign 2 solicitation, which comprise both secure cluster members and less secure candidates. We later noted aperiodically variable stars among a number of other programs targeting cool dwarfs and therefore added objects from programs GO2104, GO2051, GO2069 GO2029, GO2106, GO2089, GO2092, GO2049, GO2045, GO2107, GO2075, and GO2114 if they also had proper motions consistent with Upper Sco. \citet{Cody17}, cull some of the less confident young star candidates using {\em WISE} photometry to eliminate giant stars and other contaminants. This vetting results in a reduced set of 1443 $\rho$ Oph and Upper Sco candidate members.

For each star in this set, we downloaded the target pixel file (TPF) from the Mikulski Archive for Space Telescopes (MAST). Each target is stored under its Ecliptic Plane Input Catalog (EPIC) identification number, as listed in Table~1. Data for each object includes 3811 $\sim$10$\times$12 pixel stamp images, obtained between 2014 23 August and 10 November.

Since the loss of a second reaction wheel during the {\em Kepler} mission in May of 2013, telescope pointing for $K2$ has suffered reduced stability and requires corrective thruster firings approximately every six hours. We find a corresponding target centroid drift at a rate of $\sim$0.1\arcsec, or $\sim$0.02 pixels, per hour. While this movement is relatively small, associated detector sensitivity variations at the few percent level per pixel compromise the otherwise exquisite photometry, introducings jumps in measured flux on the same six-hour timescales.

It is helpful to track the $x$-$y$ position drift for each star over time, as this can be used for aperture placement and later detrending of the light curves. TPF headers provide a rough world coordinate system solution which is the same for all images, but these are not precise enough to center the target. We therefore cut out a 5$\times$5 pixel region around the specified target position, and used this to calculate a flux-weighted centroid. 

We carried out photomety with moving apertures, the centers of each specified by the measured centroid locations. This approach helps to minimize the effect of detector drift on the photometry.  Circular apertures were used with radii ranging from 1.0 to 4.0 pixels, in intervals of 0.5 pixels. We found that photometric noise levels after detrending for position jump effects were generally minimized with the 2-pixel aperture, though for a few objects we selected the 1.5 or 3-pixel apertures. These sizes have the additional advantage of being small enough so as to avoid flux contamination from other stars lying $\sim$12\arcsec\ away.

To clean the data, we discarded the first 93 light curve points, for which the pixel positions were particularly errant compared to the rest of the time series. We also removed points with detector anomaly flags. Finally, we pruned points lying more than five standard deviations off the median light curve trend. This was accomplished by median smoothing on $\sim$2-day timescales, removing outliers, and then adding the median trend back in.

In general, our raw moving aperture photometry consists of lower levels of pointing-related systematic jitter than for the fixed aperture case. For all of the stars discussed in this work, the amplitude of intrinsic variability dwarfs these systematics, and no further corrections are needed. However, this is not the case for the less variable cluster members which form a control population for comparison of variability demographics. In these light curves, a prominent sawtooth pattern appears on the $\sim$6-hour timescales corresponding to thruster firings, an effect that was mitigated with the detrending procedure described in \citet{2016MNRAS.459.2408A}, as described in \citet{Cody17}.

In one exceptional case (EPIC~203954898/2MASS J16263682-2415518), the light of a highly variable star was contaminated by a close neighbor 8\arcsec\ away. Single pixel photometry shows that this object undergoes high amplitude bursts, while the neighboring star is relatively constant and slightly fainter. Restricting the aperture to encompass only EPIC~203954898 results in degraded precision. We therefore used a 3-pixel aperture encircling both stars, and then removed the average flux of the companion by scaling the data to match the amplitude of the bursts in the single pixel light curves (the lower precision here affects only the detailed light curve morphology and not its overall amplitude). The result is a light curve with maximum burst amplitude of nearly eight times the quiescent flux level.

\subsection{New spectra and compiled spectroscopic data}

We collected both archival and new spectroscopic data at high dispersion with the Keck/HIRES spectrograph \citep{1994SPIE.2198..362V}. The new observations were obtained on one of: 2016 May 17 and 20, 2015 June 1 and 2, or 2013 June 4, UT, and covered the spectral range $\sim$4800~\AA\ to 9200~\AA\ at resolution $R\approx 36,000$.  The images were processed and the spectra were extracted and calibrated using the $makee$ software written by Tom Barlow. H$\alpha$ emission line strengths and spectral type estimates are tabulated in Table~\ref{tab:obstable}; other emission lines that were observed in these high dispersion data are discussed in the Appendix notes on individual stars. Table~1 as well as the notes section includes literature information in addition to our spectroscopic findings.

\subsection{Speckle imaging}
We obtained high-resolution speckle imaging for six of our young stars to assess multiplicity properties.
Our observations used the Differential Speckle Survey Instrument \citep[DSSI;][]{2011AJ....141...45H} on the Gemini-South telescope in 2016 June. Speckle observations were simultaneously made in two medium band filters with central wavelengths and bandpass FWHM values of ($\lambda_c$, $\delta\lambda$) = (692,47) and (883,54)~nm. Each star was observed for approximately 10 minutes during which time we obtained 3-5 image sets consisting of 1000, 60~ms simultaneous frames. These observations were made during clear weather at airmass 1.0 to 1.3, when the native seeing was 0.4-0.6 arcsec. Details of speckle observations using the Gemini telescope and our data reduction procedures can be found in \citet{2012AJ....144..165H} and \citet{2011AJ....142...19H}.

\section{Identification of accretion bursts}

 \begin{figure*}
 \epsscale{0.60}
 \plotone{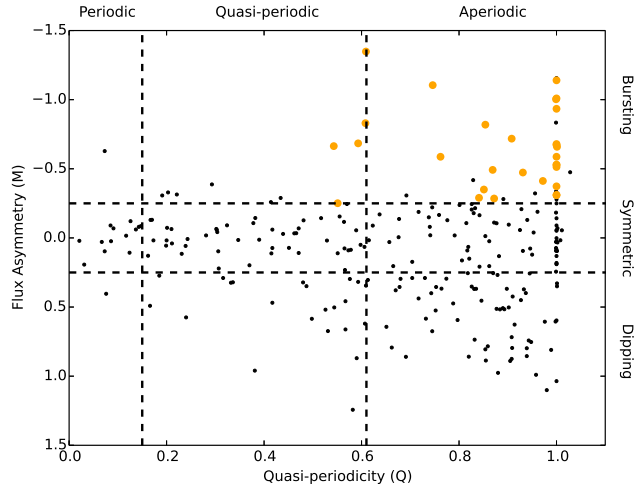}
 \caption{$Q$ and $M$ values for all stars in $\rho$ Oph and Upper Sco observed by $K2$. Black dashed lines indicate the approximate borders between different variability behaviors, as identified by \citet{2014AJ....147...82C}. Objects identified in this paper as bursters are highlighted in orange; all lie above the $M=0.25$ boundary.}
 \label{mqplot}
 \end{figure*}

We conducted visual examination of all 1443 young star light curves in order to identify stars in a ``bursting" state. Such objects were selected by identifying behavior consistent with that presented in \citet{2014AJ....147...82C} and \citet{2014AJ....147...83S}. To confirm our visually identified bursters, we also computed quantitative variability metrics, specifically the $M$ and $Q$ statistics defined by us in the papers above. $M$ describes the degree of symmetry of the light curve about its mean value. It is calculated by determining the ratio of the mean of magnitude data in the top and bottom deciles to the median of all light curve points. $M$ achieves negative values when there is a significant number of points brighter than the median, but not so many faint points.  
Unlike what was done in \citet{2014AJ....147...82C}, we calculated the $M$ statistic from flux, rather than magnitude values. The main effect of this change is to lower $M$ values (by $\sim$10\% on average), since the magnitude to flux conversion makes bright peaks more pronounced. We argue that flux units are a more natural choice here, as flux correlates with luminosity and accretion rate. Bursting light curves are highly asymmetric with frequent flux increases over the mean, resulting in $M$ values from approximately -0.3 to -1.3.  While the boundary is somewhat subjective, all objects selected by eye meet the previously defined $M<-0.25$ criterion for burster status.

The $Q$ statistic defined in \citet{2014AJ....147...82C} describes tendency towards or away from periodicity over the time series. It is a measurement of how much the standard deviation shrinks when the light curve is phased to its dominant periodicity and the associated pattern is repeated and subtracted out from the raw time series. Strictly periodic behavior (i.e., complete removal of the phase pattern) returns $Q=0$ while light curves with no repeating behavior have $Q=1$. In a few cases for which the light curves are entirely aperiodic, this removal process actually {\em increases} the underlying standard deviation; this is why the computed $Q$ value is occasionally greater than 1.0. In previous work, we have denoted light curves with moderate $Q$ values of 0.15--0.60 as ``quasi-periodic.''  We emphasize here that this range is somewhat subjective and was based on a by-eye analysis of {\em CoRoT} data on the NGC~2264 cluster. As explained in Cody et al.\ 2016, the present K2 dataset contains some objects with $Q>0.6$ that nevertheless display repeating components upon visual examination.

The $M$ and $Q$ values for the selected bursters are provided in Table~2. They are also plotted in Fig.\ \ref{mqplot}, alongside the values for other disk-bearing young stars in the $K2$ field, as selected in \citet{Cody17}. The $K2$ burster sample behavior ranges from periodic to quasi-periodic to aperiodic, with light curves most tending towards aperiodicity. Objects falling in the bursting section of the diagram but not highlighted as such tend to be long-timescale variables for which the trend removal failed. We favor our by-eye classification over the $M$ and $Q$ statistics here and thus do not consider these bursters. Several other objects with highly negative $M$ values are dominated by periodic modulation and only display zero or one bursting event. We also leave these out of the sample, in light of classification ambiguity. Overall, 18 objects with $M<-0.25$ (i.e., a value that would qualify them as busters) were removed from the sample. Most of these may be seen in Figure~\ref{mqplot} as the black points lying above the $M=-0.25$ line (apart from five that are hidden under orange burster points); the majority are only marginally above it. 

In total, we have selected 29 stars as bursters in $\rho$ Oph and Upper Sco; division into the two regions was based on a $1.2\arcdeg\times 1.2\arcdeg$ square surrounding the position RA=246.79, Dec=-24.60 to define the extent of $\rho$~Oph \citep[see][]{Cody17}. We list the basic properties of the bursters in Table~1 and show their light curves in Figure \ref{burstlcs}. The burster class exhibits several subsets of behavior, with some light curves displaying a nearly continuous series of events, and others exhibiting more discrete brightening events. We discuss the timescales of bursting in Section~6.  

\begin{turnpage}
\begin{deluxetable*}{llcccccc}
\tabletypesize{\scriptsize}
\tablecolumns{8}
\tablewidth{0pt}
\tablecaption{Young stars exhibiting bursting behavior in $K2$ Campaign 2}
\tablehead{
\colhead{EPIC id} & \colhead{2MASS id} & \colhead{Other ids} & \colhead{SpT} &  \colhead{EW H$\alpha$} & \colhead{H$\alpha$ 10\%} & \colhead{Refs} & \colhead{Region} \\ 
\colhead{} & \colhead{} & \colhead{} & \colhead{} &  \colhead{(\AA)} & \colhead{(km~s$^{-1}$)} & \colhead{} & \colhead{} \\
}
\startdata
203382255 & J16144265-2619421 & & {\bf M4-M5.5} & {\bf -77} & {\bf 154} & 1 & USco\\
203725791 & J16012902-2509069 & USco CTIO 7 & {\bf M2}/M3.5 & {\bf -170}, -129 & {\bf 437} & 1, 2 & USco\\
203786695 & J16245974-2456008 & WSB 18 & M3.5 & -8.4, -140 & - & 3 & $\rho$ Oph\\
203789507 & J15570490-2455227 &    & - & - & - & & USco\\
203794605 & J16302339-2454161 & WSB 67 & {\bf M3.5-M5} & {\bf -69} & {\bf 485} & 1 & $\rho$ Oph\\
203822485 & J16272297-2448071 & WSB 49, MHO 2111, DROXO 57 & M4.25 & -37 & - & 4 & $\rho$ Oph\\
203856109 & J16095198-2440197 &  & {\bf M5-M5.5} & {\bf -15} & {\bf 155} & 1 & USco\\
203899786 & J16252434-2429442 & V852 Oph, SR 22, DoAr 19, WSB 23 & M4.5/M3 & -31, -170 & - & 4, 5, 6 & $\rho$ Oph\\
203905576 & J16261886-2428196 & VSSG 1, Elias 20, YLW 31, ISO-Oph 24, & {\bf K7--mid-M}/M0 & {\bf -70} & {\bf 416} & 1,7 & $\rho$ Oph\\
                   &                                   & IRAS 16233-2421, MHO 2103   &     &       &  &  &\\
203905625 & J16284527-2428190 & V853 Oph, SR 13, DoAr 40, WSB 62,  & M3.75 & -30, -48, -46 & - & 4 & $\rho$ Oph\\
                   &                                   & ISO-Oph 199, HBC 266            &       &  &  & &\\
203913804 & J16275558-2426179 & V2059 Oph, DoAr 37, SR 10, ISO-Oph 187, & M2 & -43, -56, -108 & - & 4,8 & $\rho$ Oph\\
                   &                                   & WSB 57, YLW 56, HBC 265, SVS 1771                       &       &              &    & & \\
203928175 & J16282333-2422405 & SR 20W &  K5 & -35 & - & 3 & $\rho$ Oph\\
203935537 & J16255615-2420481 & V2058 Oph, DoAr 20, Elias 13, & K4.5 & -220, -87, -67 & - & 4 & $\rho$ Oph\\
                   &                                   & SR 4, WSB 25, YLW 25, IRAS 16229-2413, &         &                        &    &     & \\
                   &                                   & MHA 365-12, ISO-Oph 6      &          &                        &    &    & \\
203954898 & J16263682-2415518 & ISO-Oph 51 & M0 & -10 & - & 9 & $\rho$ Oph\\
204130613 & J16145026-2332397 & BV Sco & M4.5 & -108 & - & 10 & USco\\
204226548 & J15582981-2310077 & USco CTIO 33, USco 42  & M3  & -158, -250 & - & 11,12 & USco\\
204233955 & J16072955-2308221 &  & M3 & -150 & - & 10  & USco\\
204342099 & J16153456-2242421 & VV Sco, IRAS 16126-2235, PDS 82a & M1/{\bf K9-M0} & -20, {\bf -31} & {\bf 337} & 13, 1 & USco\\
204347422 & J16195140-2241266 &  & - & - & - &  & USco\\
204360807 & J16215741-2238180 &  & {\bf M6} & {\bf -140} & {\bf 341} & 1 & USco\\
204397408 & J16081081-2229428 &  & M5.75/M5 & -22, -49, -31  & - & 10, 14, 15 & USco\\
204440603 & J16142312-2219338 &  & M5.75 & -95 & - & 10 & USco\\
204830786 & J16075796-2040087 & IRAS 16050-2032 & M1/{\bf G6-K5} & -357, {\bf -165} & {\bf 684} & 16, 1 & USco\\
204906020 & J16070211-2019387 & KSA 68 & M5 & -8, -30 & - & 11, 17 & USco\\
204908189 & J16111330-2019029 &  & {\bf M1}/M3 & {\bf -160} & {\bf 324} & 1,18 & USco\\
205008727 & J16193570-1950426 &  & {\bf K7-M3} & {\bf -55} & {\bf 322} & 1 & USco\\
205061092 & J16145178-1935402 &  & {\bf M5-M6} & {\bf -70} & {\bf 173} & 1 & USco\\
205088645 & J16111237-1927374 &  & M5, M6 & -50, -50 & - & 12, 19  & USco\\
205156547 & J16121242-1907191 &  & {\bf M5-M6} & {\bf -15} & {\bf 127} & 1 & USco\\
\enddata
\tablecomments{\label{tab:obstable} Stars in the $K2$ Campaign 2 burster sample, in order of EPIC id. EPIC~203786695/2MASS J16245974-2456008 has a companion at 1.1\arcsec\ separation, and the two H$\alpha$ values belong to the distinct components of the system. The H$\alpha$ 10\% widths in column 6 are derived from data presented in this paper. We have highlighted in bold the other new values derived as part of this work. References: 1) this work, 2) \citet{2015MNRAS.448.2737R}, 3) \citet{1997A&A...321..220B}, 4) \citet{2005AJ....130.1733W}, 5) \citet{2007ApJ...657..338P}, 6) \citet{1998MNRAS.300..733M}, 7) \citet{2010ApJ...723.1241A}, 8) \citet{1983A&AS...53..291A}, 9) \citet{2011AJ....142..140E}, 10) \citet{2011A&A...527A..24L}, 11) \citet{2009AJ....137.4024D}, 12) \citet{2002AJ....124..404P}, 13) \citet{1998A&A...333..619P}, 14) \citet{2008ApJ...688..377S}, 15) \citet{2012ApJ...745...56D}, 16) \citet{2009ApJ...703.1511K}, 17) \citet{2001AJ....121.1040P}, 18) \citet{2012ApJ...758...31L}, 19) \citet{2010A&A...517A..53M}. 
}
\end{deluxetable*}
\end{turnpage}

 \begin{figure*}
 \epsscale{0.90}
 \plotone{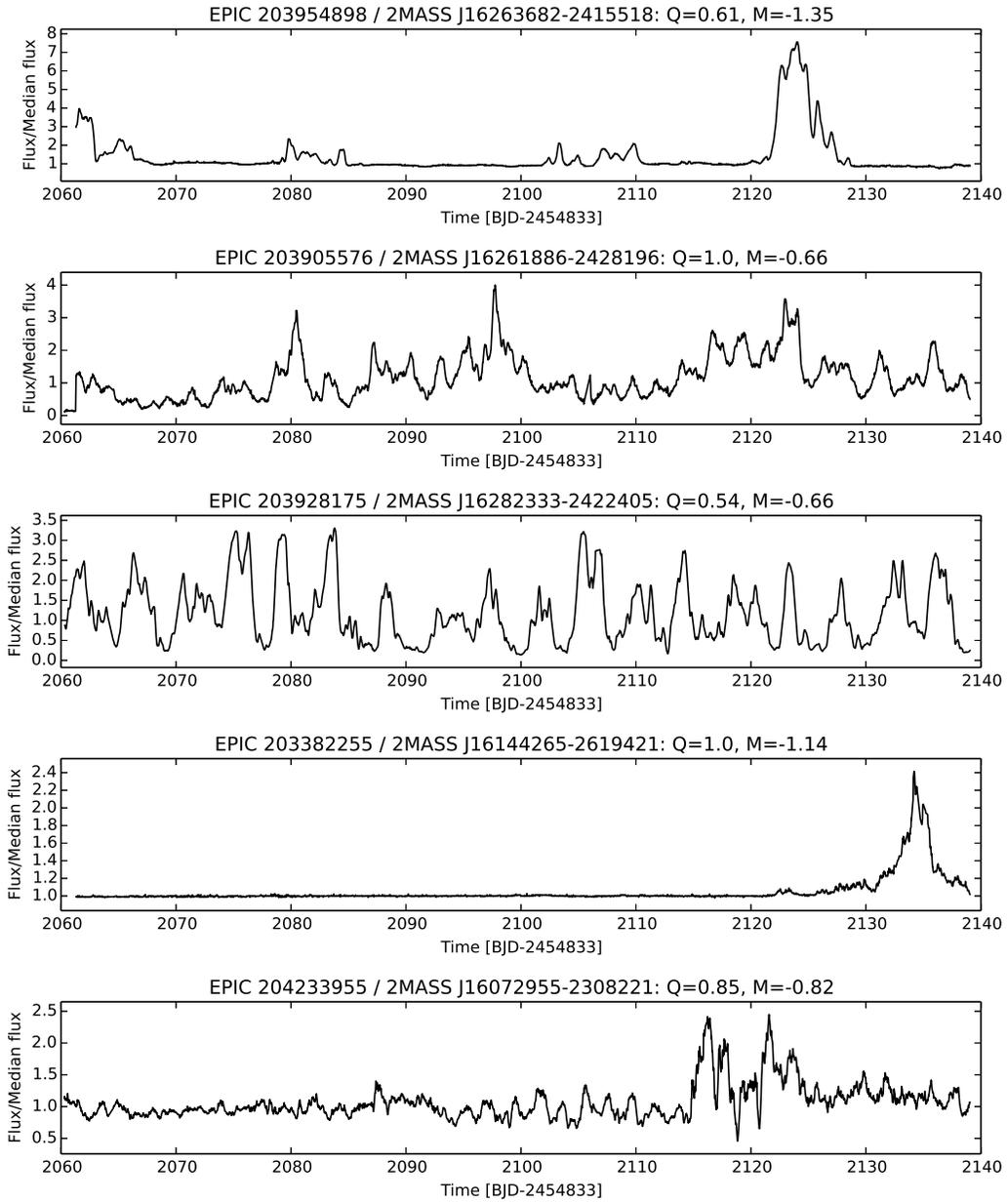}
 \caption{Light curves of selected bursters over the 80-day duration of $K2$ Campaign 2, in approximate order of amplitude. 
 }
 \label{burstlcs}
 \end{figure*}
 
 \addtocounter{figure}{-1} 
 \begin{figure*}
 \epsscale{0.90}
 \plotone{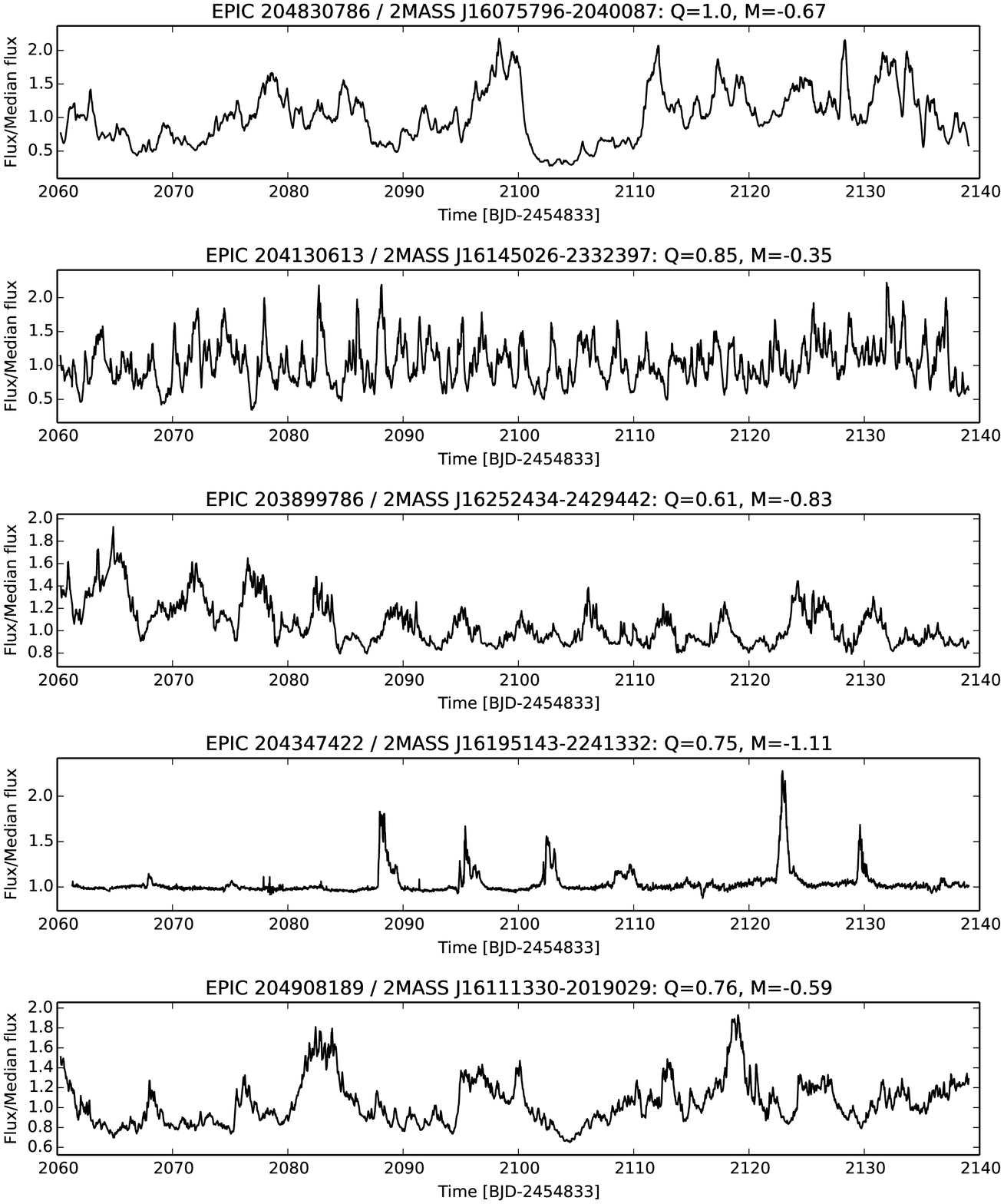}
 \caption{Cont.}
  \end{figure*}
  
 \addtocounter{figure}{-1} 
 \begin{figure*}
 \epsscale{0.90}
 \plotone{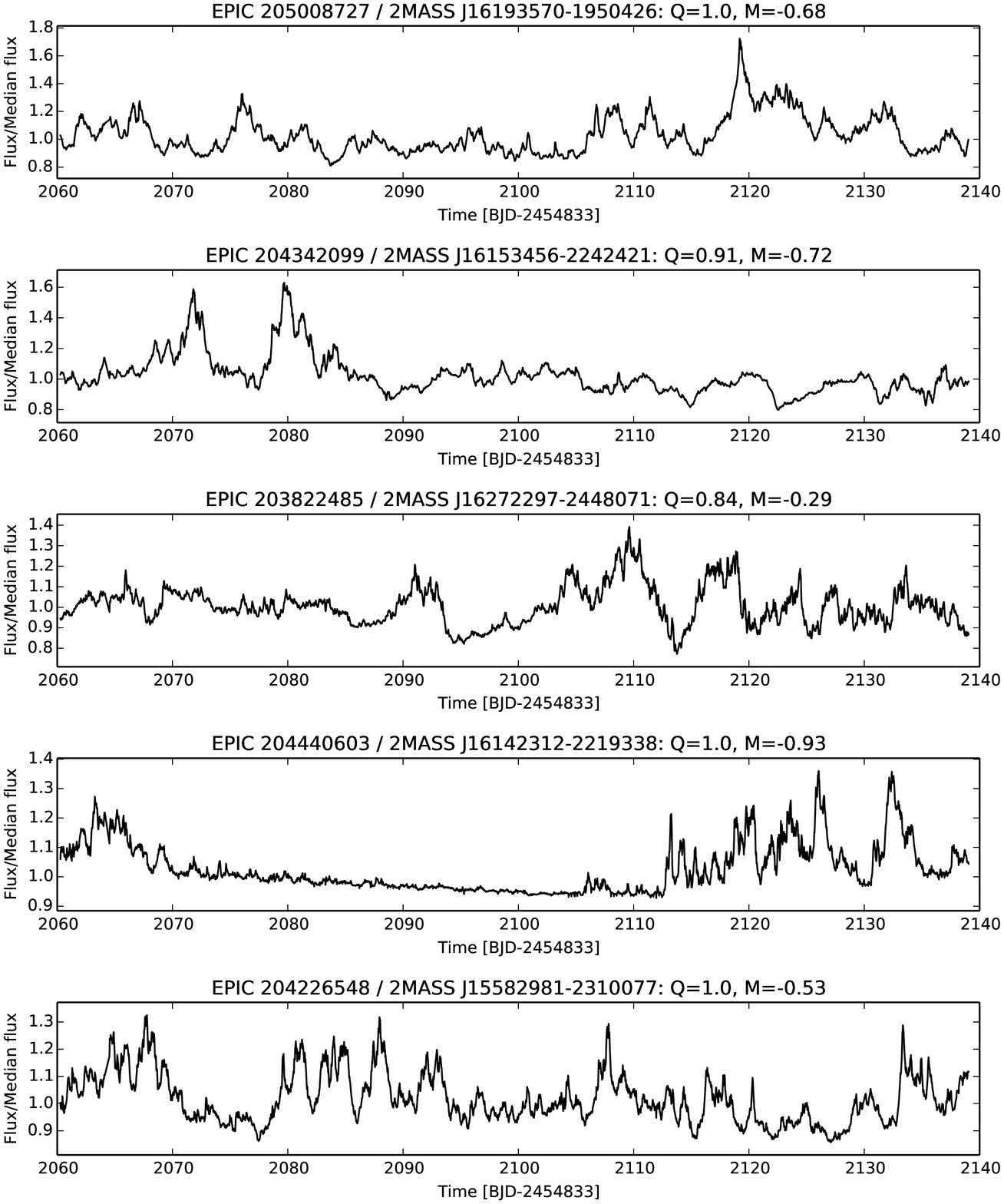}
 \caption{Cont.}
  \end{figure*}
  
  \addtocounter{figure}{-1}
  \begin{figure*}
 \epsscale{0.90}
 \plotone{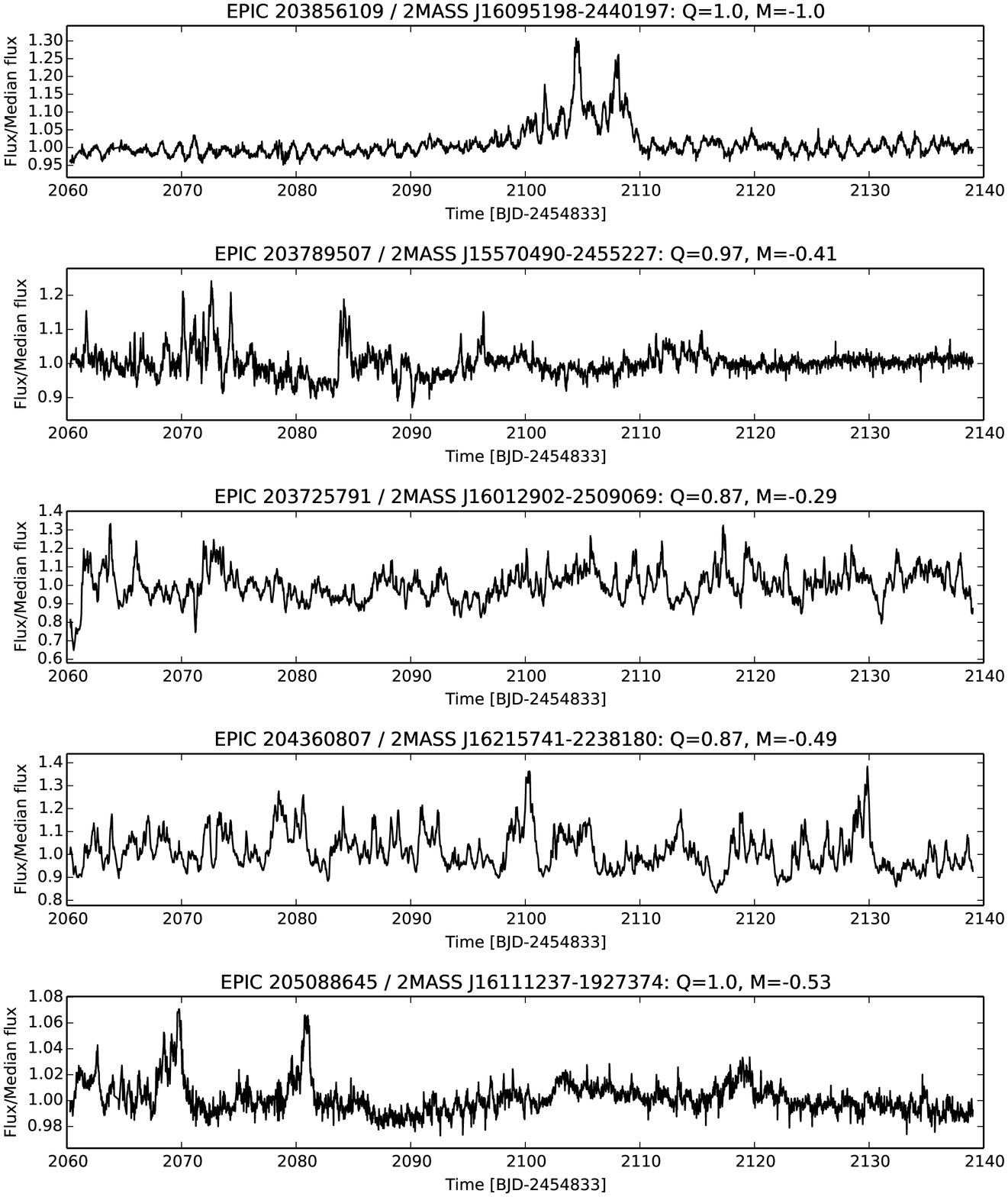}
 \caption{Cont.}
  \end{figure*}
  
  \addtocounter{figure}{-1}
  \begin{figure*}
 \epsscale{0.90}
 \plotone{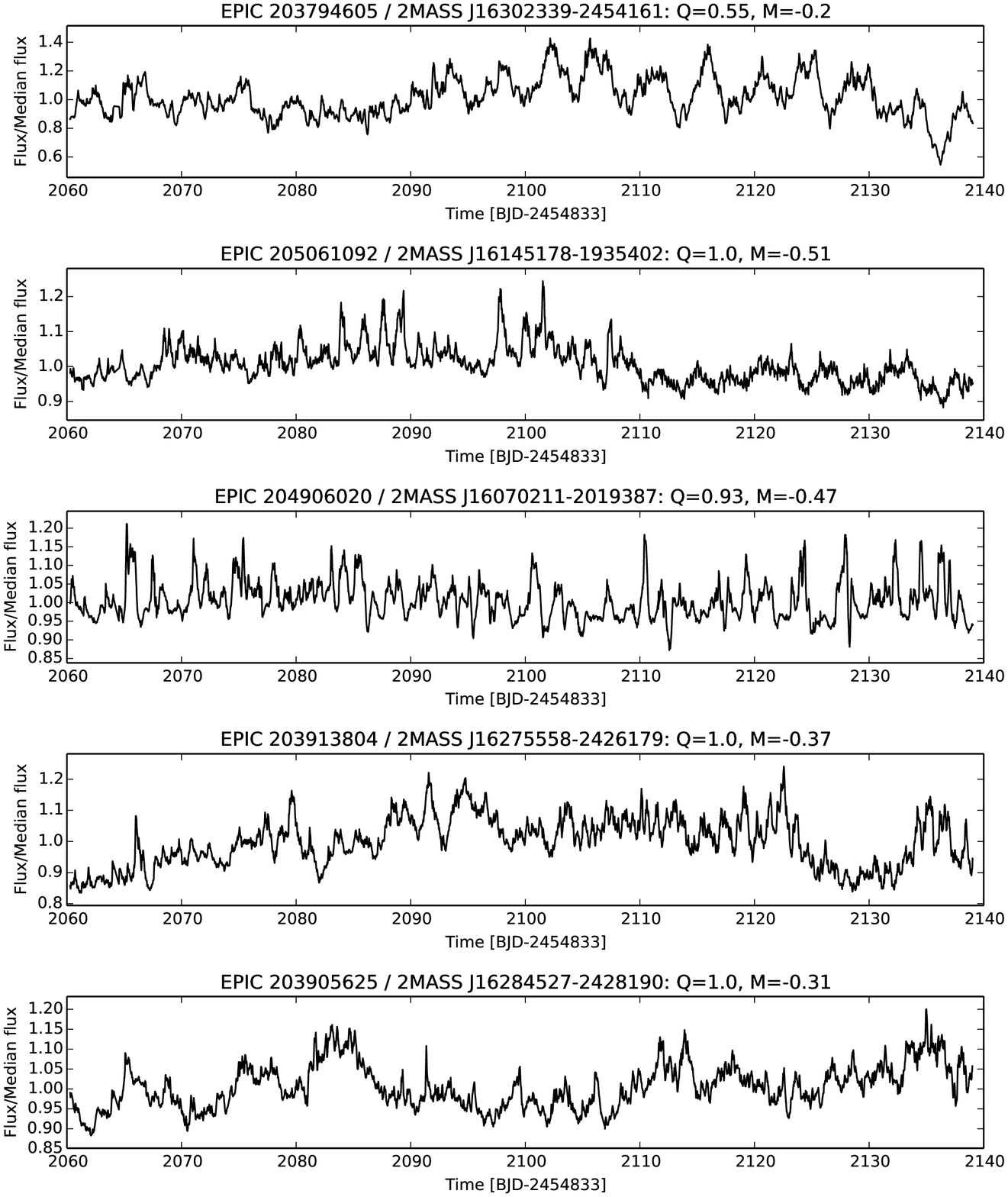}
 \caption{Cont.}
  \end{figure*}

  \addtocounter{figure}{-1}
  \begin{figure*}
 \epsscale{0.90}
 \plotone{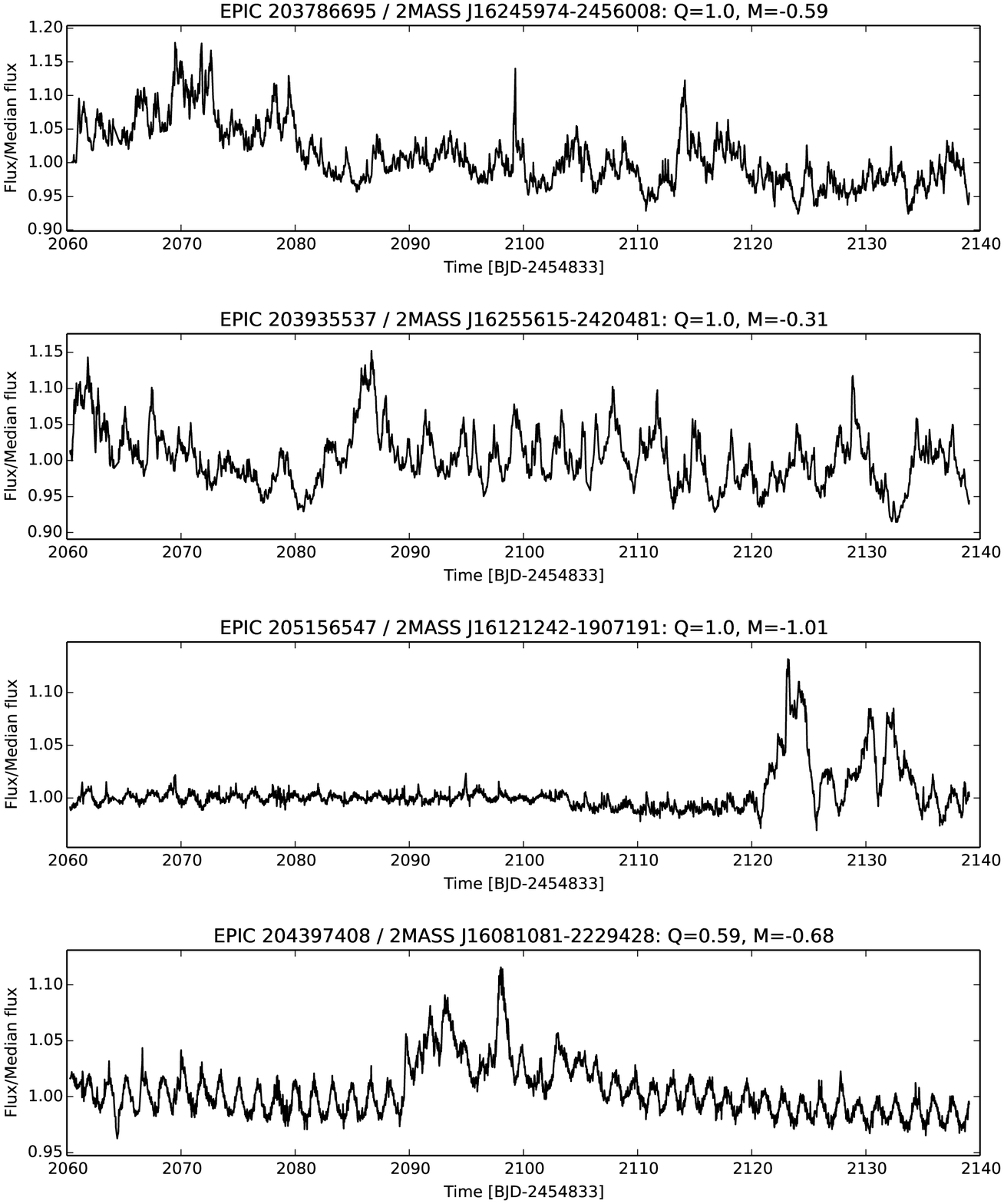}
 \caption{Cont.}
  \end{figure*}

\section{Disk and accretion properties of the burst sample}

A hypothesis for the short-lived, often repeatable, brightening events in our 29 $K2$ light curves is that they are caused by episodic accretion from the circumstellar disk onto the young star. A specific requirement for the accretion-driven burst hypothesis is that the objects exhibit both infrared excess indicative of circumstellar dust, serving as the reservoir for the accretion, and either ultraviolet excess or line emission from hot gas in the nearby circumstellar environment. The dust criterion is satisfied by our burster sample, as all stars have been already selected as infrared excess sources in \citet{Cody17}. As shown in that work, there are 344 disk bearing stars in the entire $K2$ Campaign 2 sample, of which 299 are bright enough to obtain light curves. This number includes all 29 burster stars identified here. Therefore, the fraction of bursters among the total disk-bearing sample is at least 8$\pm$2\%. The error comes from consideration of the Poisson uncertainties.
These values can also be considered separately for the $\rho$~Oph (128 disked stars; 92 with light curves) and Upper Sco (216 disked stars; 207 with light curves) samples. There are 10 bursters in $\rho$~Oph and 19 bursters in Upper Sco; both numbers lead to roughly the same value of 8--10\%, with a $\sim$2\% error on this fraction. 

\subsection{Circumstellar Dust}

Infrared color-magnitude diagrams can also shed light on what photometric aspects, if any, separate burster stars from other disk bearing sources. We show the color-color diagrams $J-K$ versus $K-W3$ and $J-K$ versus $K-W4$ in Figure \ref{JKWISE} and the spectral energy distributions in Appendix Figure \ref{seds}, to illustrate the strength of emission at farther, cooler locations in the disk. The vast majority of the bursters have excesses in the three longest wavelength WISE bands $W2$, $W3$, and $W4$ ($4.6 \mu$m, $12 \mu$m, and $22 \mu$m, respectively). The individual spectral energy distributions in the Appendix (Fig.~\ref{seds}) provide a finer look at the circumstellar flux patterns. This finding is in stark contrast to the overall disk sample, in which only 50\% of objects (or a total of 172) have excesses in all three bands. Thus, the presence of a full, minimally evolved disk appears to be preferred for bursting behavior. 

\begin{figure*}
 \epsscale{1.0}
 \plotone{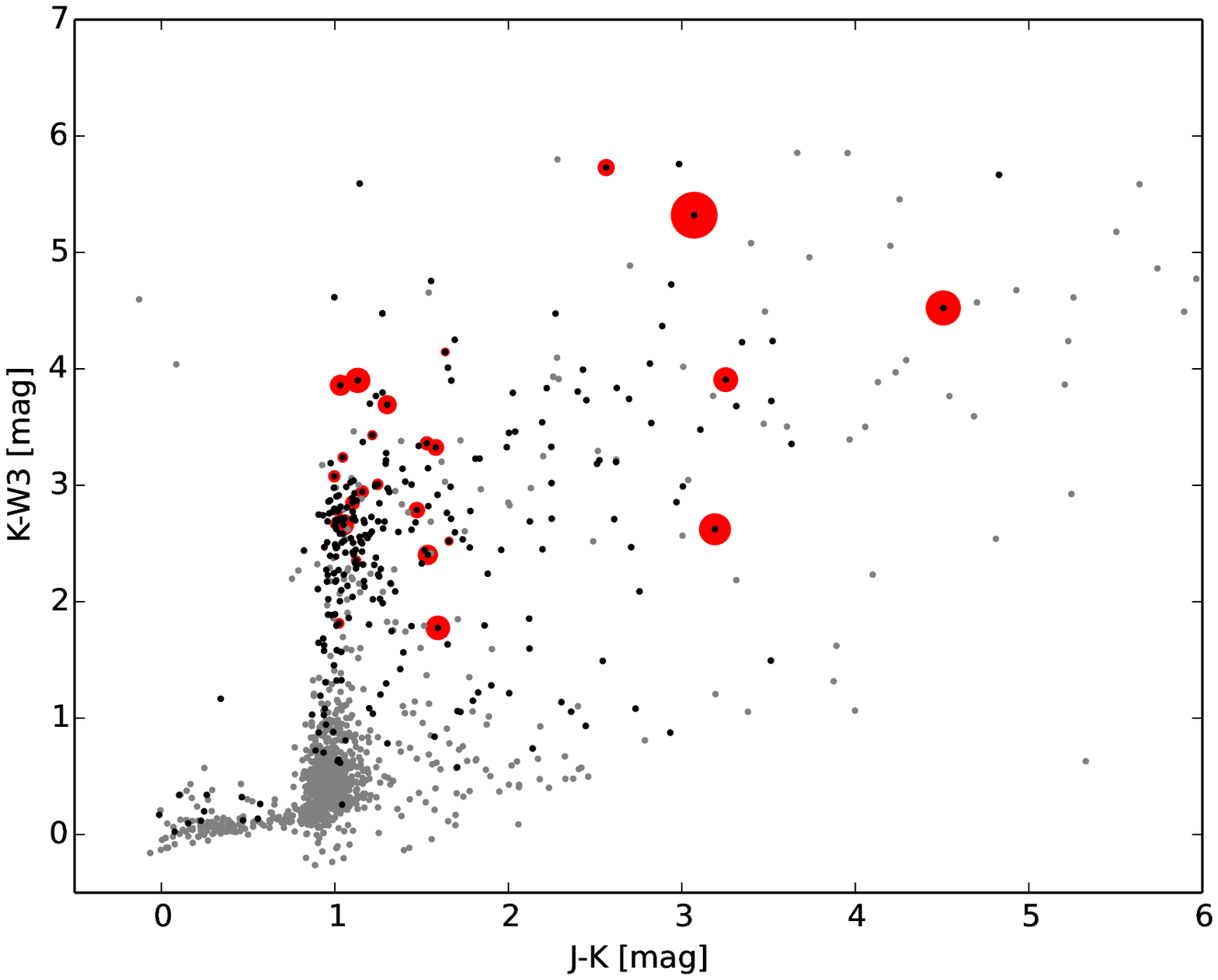}
 \plotone{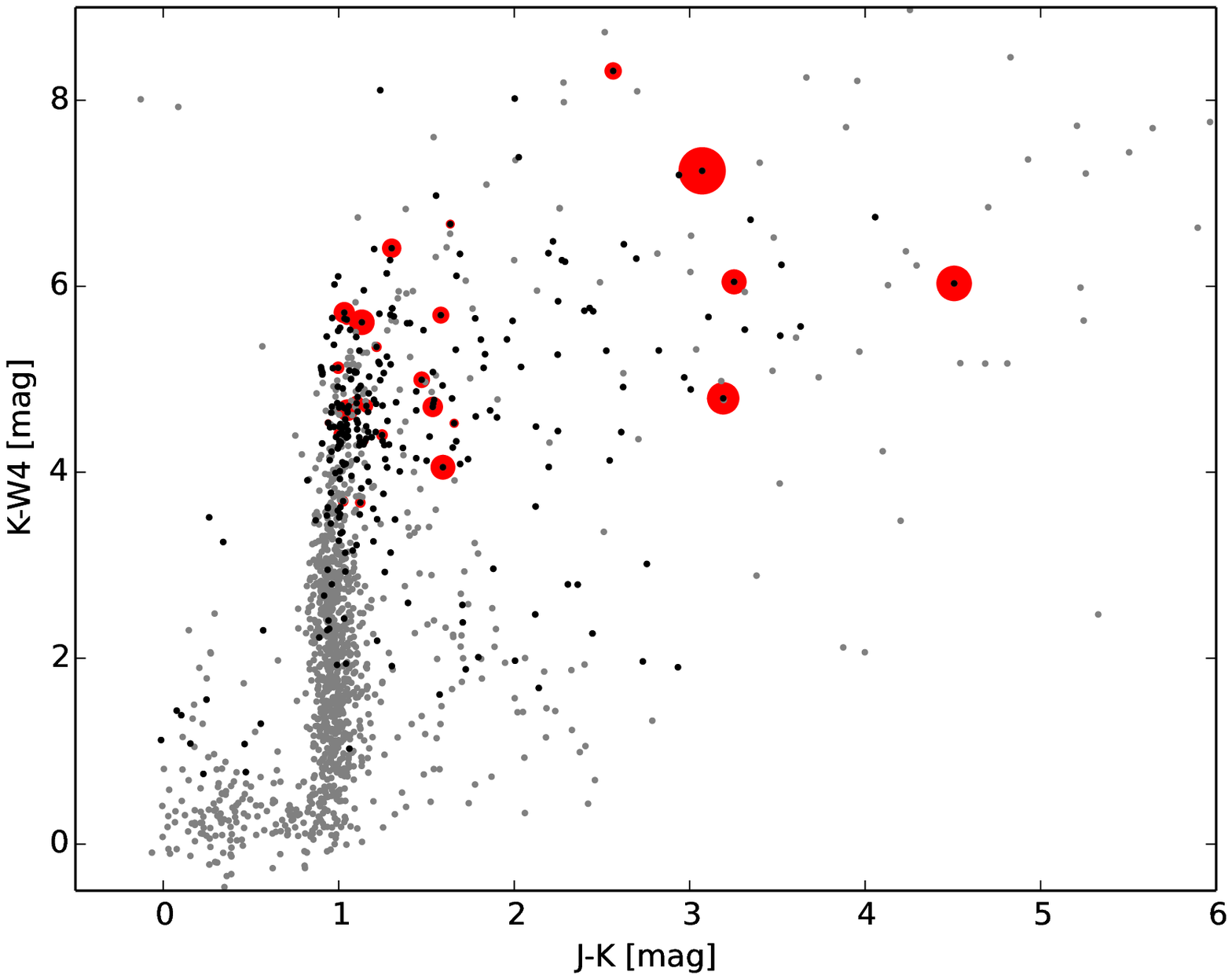}
 \caption{Near and mid-infrared color-color diagram of stars in the Upper Sco/$\rho$ Oph regions (grey), with disk-bearing stars in black and bursters highlighted in red. Burster point sizes are scaled by light curve amplitude.}
 \label{JKWISE}
 \end{figure*}

It has traditionally been thought that young stars with prominent fading events are surrounded by nearly edge-on disks; the dust clumps routinely obscure the central star \citep[e.g.][]{2003A&A...409..169B,2013A&A...557A..77B}, causing decrements in the light curve. Conversely, one might imagine that for disk systems farther from edge-on orientation, we would have a more direct view of the accretion columns and shocks near the stellar poles. This could allow relatively unfettered observation of accretion bursts.  Looking at the selection of 172 full disks identified in \citet{Cody17}, 90\% are variable, but only 17\% are bursting. If this is a reflection of geometric selection, then one might hypothesize that burster disks are viewed at angles ranging from face-on to $\sim$34\arcdeg. However, this assumes that face-on is the best angle at which to view bursting; as suggested by the ALMA data described below, this may not be the case.

The idea that bursting behavior may be a function of viewing angle can be further explored by considering resolved disk imaging. Five of the bursters discussed in this paper have been observed with ALMA at 0.88~mm \citep{2014ApJ...787...42C,Baren16}.  EPIC~204830786 (2MASS~J16075796-2040087) has a broad CO $J=3-2$ line detection, with velocities from -17 to 17 km~s$^{-1}$; the profile suggests that the disk is {\em not} face-on. EPIC~204342099 (2MASS~J16153456-2242421) is the only other disk with a CO detection, albeit a weak one. The broad velocity distribution again suggests that this system is not oriented face-on. The other three disk-bearing sources in our sample (EPIC~204906020/2MASS~J16070211-2019387, EPIC~204226548/2MASS~J15582981-2310077, EPIC~204908189/2MASS J16111330-2019029) have no CO detection. The latter two do show continuum, so CO may be highly depleted or the disks in these cases could be physically small. 

The continuum fluxes can also be used to infer disk properties. \citet{Baren17} infers from modeling SEDs and size constraints that EPIC~204830786 (2MASS~J16075796-2040087) has inclination 42\arcdeg$^{+12}_{-9}$, while EPIC~204342099 (2MASS~J16153456-2242421) is inclined at 43\arcdeg$^{+15}_{-16}$. Dust masses were inferred by \citet{Baren16} and range from $<$0.5~$M_\oplus$ (EPIC~204906020/2MASS~J16070211-2019387; non-detection) to 9.3~$M_\oplus$ in the case of EPIC~204830786/2MASS~J16075796-2040087. All five bursters in the ALMA sample are noted as having been classified as full disks (as opposed to evolved or transitional) by \citet{2012ApJ...758...31L}. 

These intermediate inclination values are consistent with the idea proposed above that we are not looking through disk material, which would be expected to produce ``dipping" rather than ``bursting'' light curves.  But they are inconclusive regarding whether we have a direct view of material accreting onto the central star.

 \subsection{Gas and Accretion}

\begin{figure}
 \epsscale{1.25}
 \plotone{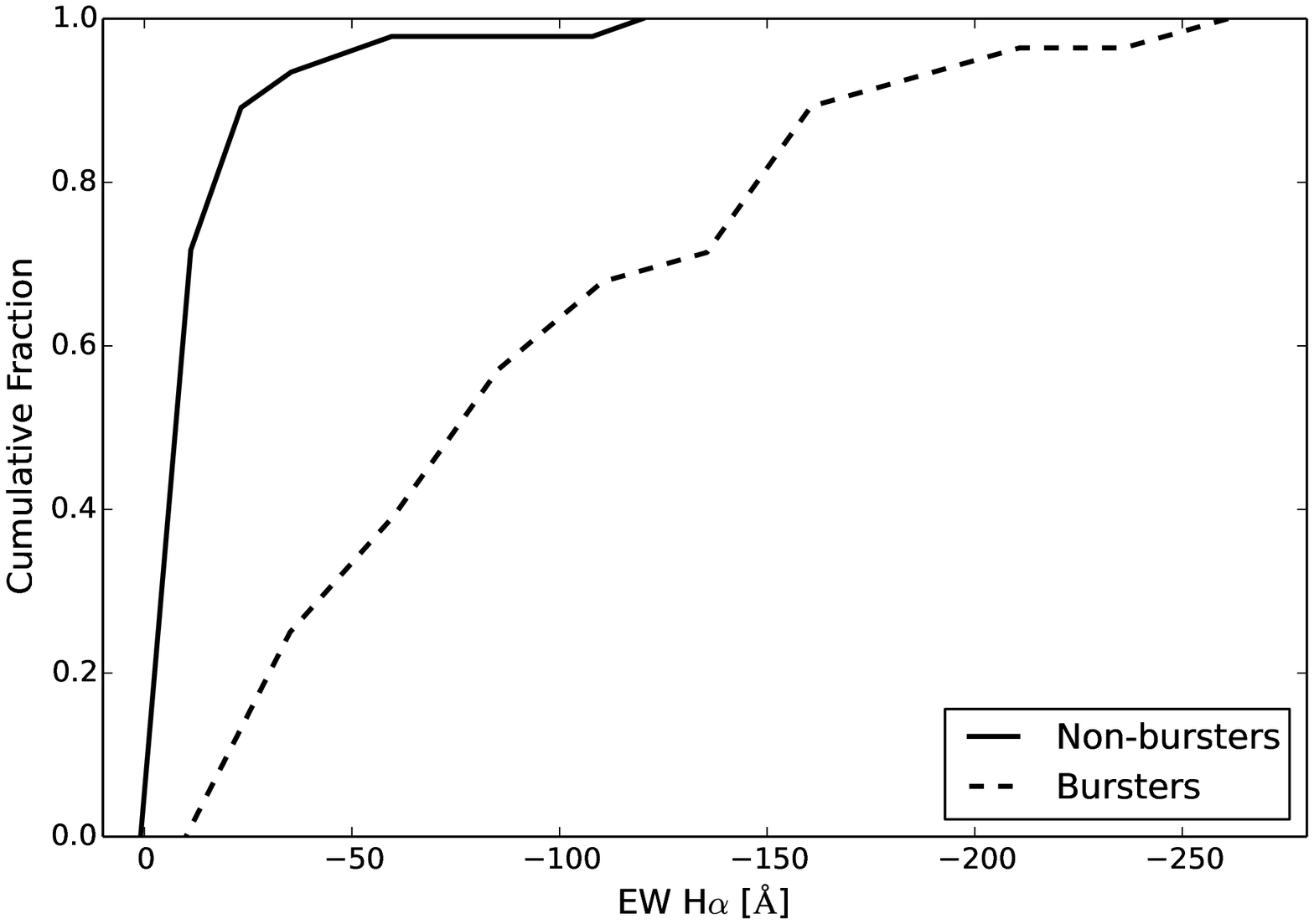}
 \caption{We show the cumulative distributions of H$\alpha$ equivalent width for disk-bearing stars in our K2 young star set with available spectroscopy; this set includes 28 bursters and 46 non-bursters. Where more than one H$\alpha$ measurement is available, we adopt the mean value. The sample was binned in 10\AA\ wide sets to produce the distribution. While the non-bursting stars tend to predominate between 0 and -15\AA\ (i.e., emission), the burster H$\alpha$ values show a much wider dispersion, reaching much values up to -250\AA. Thus the burst phenomenon seems to favor stars with high accretion rates. We note that there is a single disk-bearing star with quasi-periodic light curve that has a reported H$\alpha$ equivalent width around -120\AA. Such a value is highly unusual for a star with a non-bursting light curve.}
 \label{halpha}
 \end{figure}

In addition to infrared disk indicators, we obtained spectroscopic data from both the literature and our own high-resolution follow-up spectroscopy (\S2.2). The H$\alpha$ emission equivalent width (EW) and 10\% width values (Table~1) are indicative of significant accretion. As a control sample, we gathered H$\alpha$ EWs for other disk-bearing (but not necessarily bursting) stars in Upper Sco from \citet{2015MNRAS.448.2737R}, \citet{2009AJ....137.4024D}, and \citet{2002AJ....124..404P}. We cross-matched them against our Upper Sco/$\rho$~Oph K2 $WISE$ excess star list and eliminated any objects not in common. We then compared the H$\alpha$ values of this general Upper Sco sample with those of the bursters in Figure \ref{halpha}. Since some of the bursters have multiple H$\alpha$ measurements (see Table~1), we have taken the average of all available values. We find that the bursters occupy a large range of H$\alpha$ EW values, from -10\AA\ (presumably a low-state value) to several hundred angstroms in emission. The non-burster stars, on the other hand, display weak H$\alpha$, with EWs primarily from 0 to -15\AA\ and a tail out to -50\AA\ with one value around -120\AA.

\begin{figure}
\epsscale{1.20}
\plotone{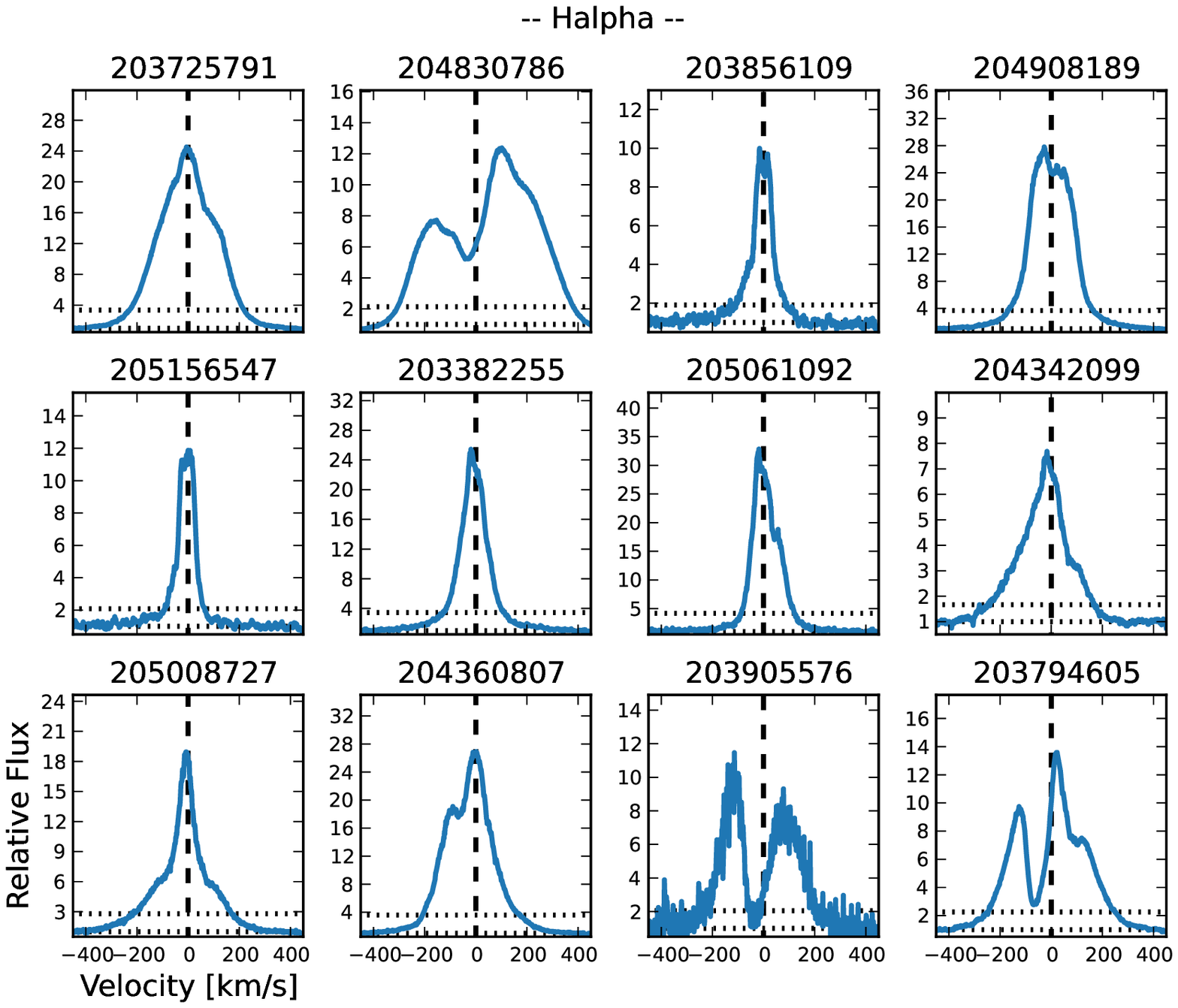}
\plotone{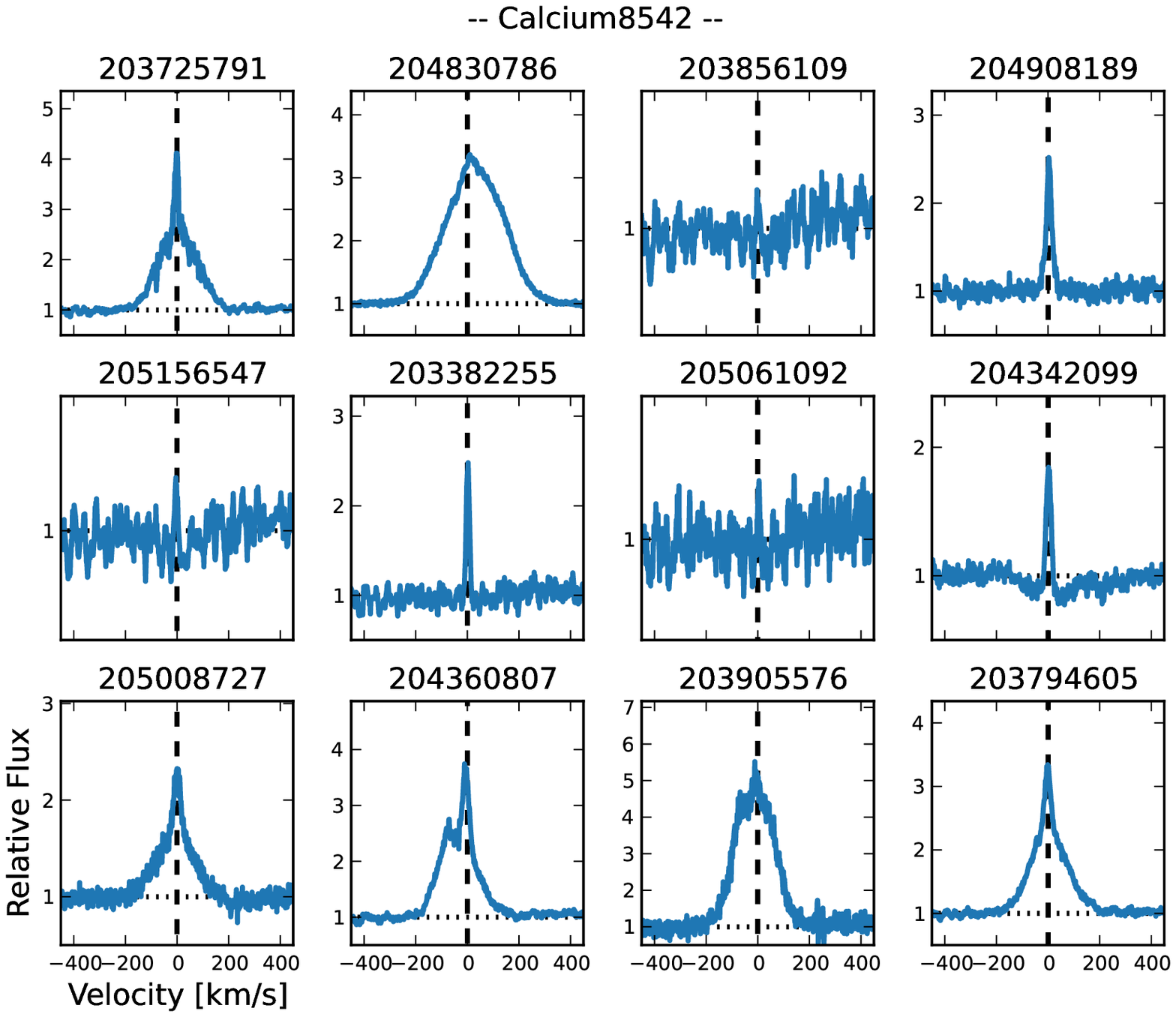}
\plotone{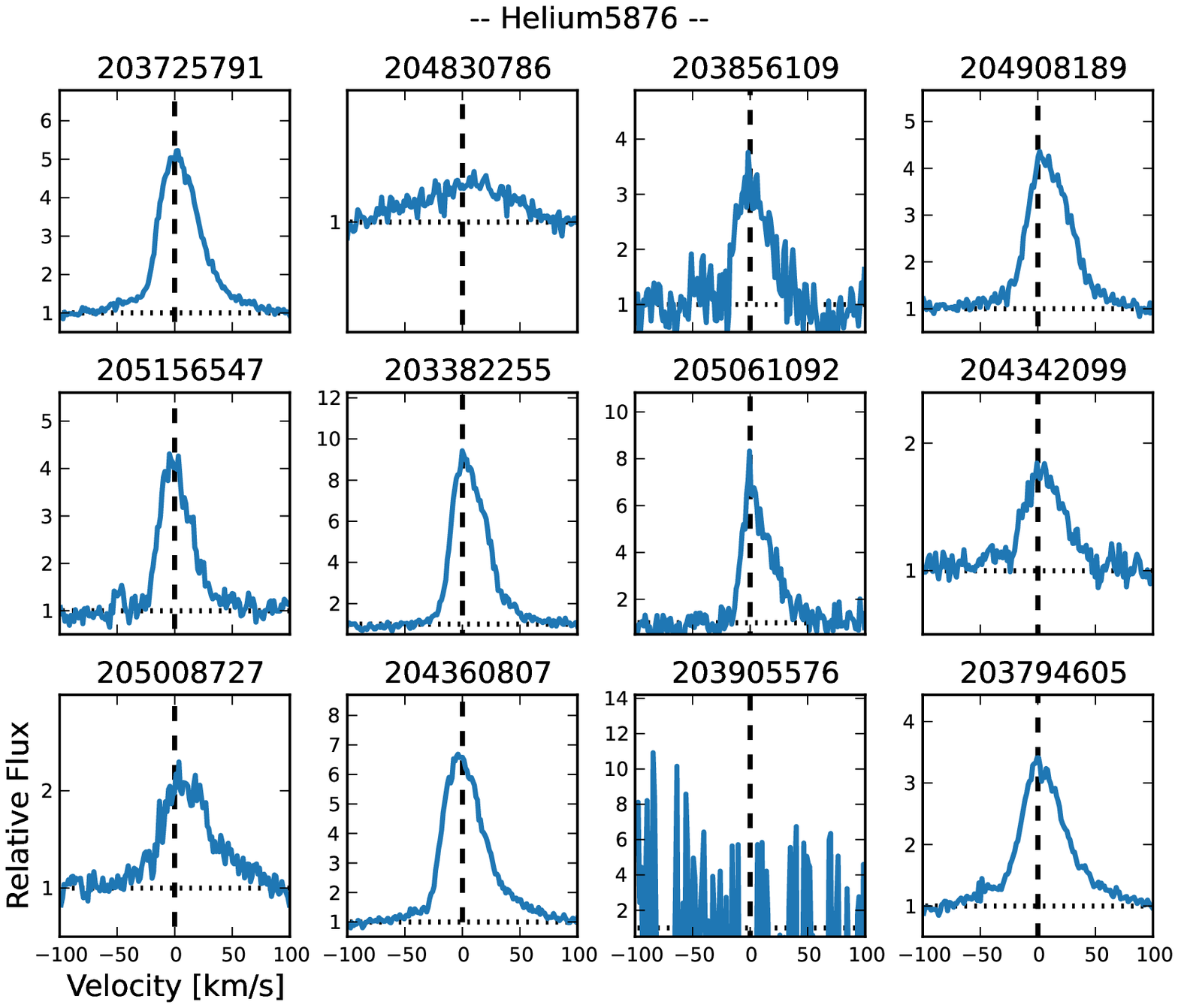}
\caption{Line profiles in H$\alpha$, \ion{Ca}{2} 8542 \AA, and \ion{He}{1} 5876 \AA\ measured by Keck/HIRES for 12 of the 29 bursters, labeled by EPIC identifier.  The H$\alpha$ panels include a horizontal line at 10\% of the peak (un-normalized) flux in addition to the horizontal line indicating the continuum level. Note the change in velocity scale for the helium panels.  Broad width and structured velocity profiles indicate accretion and wind phenomena.}
\label{profiles_halpha}
\end{figure}

Figure~\ref{profiles_halpha} illustrates the emission line profiles for the 12 out of the 29 bursters for which we have high resolution spectra in H$\alpha$, \ion{Ca}{2} 8542\AA\ and \ion{He}{1} 5876\AA. \cite{2003ApJ...582.1109W} advocated the designation of accreting stars based on H$\alpha$ line widths at 10\% of the maximum line strength that are larger than 270~km~s$^{-1}$. Although all 12 sources have velocities larger than 100~km~s$^{-1}$ (see Table~1), only slightly more than 1/2 meet the \cite{2003ApJ...582.1109W} requirement. There is a tendency for the narrower velocity stars (EPIC 203856109/2MASS~J16095198-2440197, EPIC~205156547/2MASS~16121242-1907191, EPIC~203382255/2MASS~J16144265-2619421, EPIC~205061092/2MASS~J16145178-1935402, EPIC~205008727/2MASS~J16193570-1950426) to have lower duty cycles in their burst patterns (Figure~\ref{burstlcs}) (Where available, H$\alpha$ velocity and duty cycle are correlated at a significance level of 1.2$\times 10^{-3}$). It may be that our spectra were taken at non-burst epochs. This is speculative at best, however. 

The weak H$\alpha$ sources all exhibit only narrow component emission in their
\ion{Ca}{2} profiles, as does EPIC 204342099 (2MASS~J16153456-2242421) which has a broad H$\alpha$ profile.  Generally, the broad H$\alpha$ sources exhibit both broad component and narrow component \ion{Ca}{2}.  \cite{2006A&A...456..225A} review the classic literature on \ion{Ca}{2} profile morphology in young stars, and discuss magnetospheric models of it. Our profiles do not seem to exhibit the asymmetries predicted by these models (e..g blueshifted peaks, redshifted depressions), however.  Notably, our broad-lined \ion{Ca}{2} sources also exhibit evidence for forbidden line emission, e.g. [\ion{O}{1}] 6300\AA.

The morphology of \ion{He}{1} emission lines in young stars was studied by \cite{2001ApJ...551.1037B} who also designated narrow line, broad line, and narrow$+$broad profile categories.  Our stars are dominated by their narrow component emission, with widths ranging between 40-70~km~s$^{-1}$, but some may also have weak broad components that would require line decomposition to characterize. Most of the \ion{He}{1} profiles appear to have slight asymmetries, however, in the sense of broader redshifted emission than blueshifted emission with the line peaks at zero-velocity.

Beyond emission line morphology, accretion rates are estimated for a handful of our burster stars by \citet{2006A&A...452..245N}. Values range from 10$^{-9.9}$ to 10$^{-6.7}$~$M_\odot$~yr$^{-1}$-- a large range. It should be noted that neither the H$\alpha$ EWs presented above nor the accretion rates referenced here were necessarily measured during a time when the stars were undergoing bursting events. Thus it is plausible that accretion is preferentially high in these sources, but only at certain times.

Under the hypothesis that burst events in light curves are due to higher than average mass flow, we can convert the flux to a quantitative increase in the mass accretion rate. This requires several assumptions. First, we assign to all stars in our sample a low-level baseline accretion rate, $\dot{M}_{\rm low}$, which then increases to a larger rate $\dot{M}_{\rm high}$ during bursts. We assume that this increase in mass flow can be equated to the ratio of the accretion flux $F_{\rm acc}$ in and out of the burst state through the accretion luminosity, $L_{\rm acc}$:
\begin{equation}
\frac{\dot{M}_{\rm high}}{\dot{M}_{\rm low}} = \frac{L_{\rm acc,high}}{L_{\rm acc,low}}\sim\frac{F_{\rm acc,high}}{F_{\rm acc,low}}.
\end{equation}
The accretion luminosity is related to the stellar mass $M_*$ and radius $R_*$ by $L_{\rm acc} = 1.25(GM_*\dot{M}/R_*)$ where the pre-factor is that appropriate for an assumed magnetospheric accretion scenario.

The measured flux density $F$ (i.e., in the $Kepler$ band) contains contributions from both accretion and the underlying stellar luminosity, $L_*$. We label the measured flux ratio ``$r$'':
\begin{equation}
r \equiv \frac{F_{\rm high}}{F_{\rm low}} = \frac{F_*+F_{\rm acc,high}}{F_*+F_{\rm acc,low}}
\end{equation}

$F_{\rm low}$ and $F_{\rm high}$ are approximately the minimum and maximum flux values respectively attained in a given light curve. To determine $F_{\rm acc,high}/F_{\rm acc,low}$ and hence $\dot{M}_{\rm high}/\dot{M}_{\rm low}$, we must estimate the ratio of stellar to accretion luminosity. This quantity has been studied by e.g. \citet{2014A&A...569A...5N} and we adopt the fit based on their Figure 2. For each star in the burster sample, we estimate the stellar luminosity by considering the $J$-band magnitudes of similar spectral type non-accreting young stars in the $K2$ Campaign 2 set, and applying bolometric and extinction corrections as outlined in \citet{2006A&A...452..245N}. Combining this with Equations 1 and 2, it can be shown that
\begin{equation}
\frac{\dot{M}_{\rm high}}{\dot{M}_{\rm low}}\sim\frac{L_*}{L_{\rm acc}}(r-1)+r.
\end{equation}

\begin{figure}
\epsscale{1.20}
\plotone{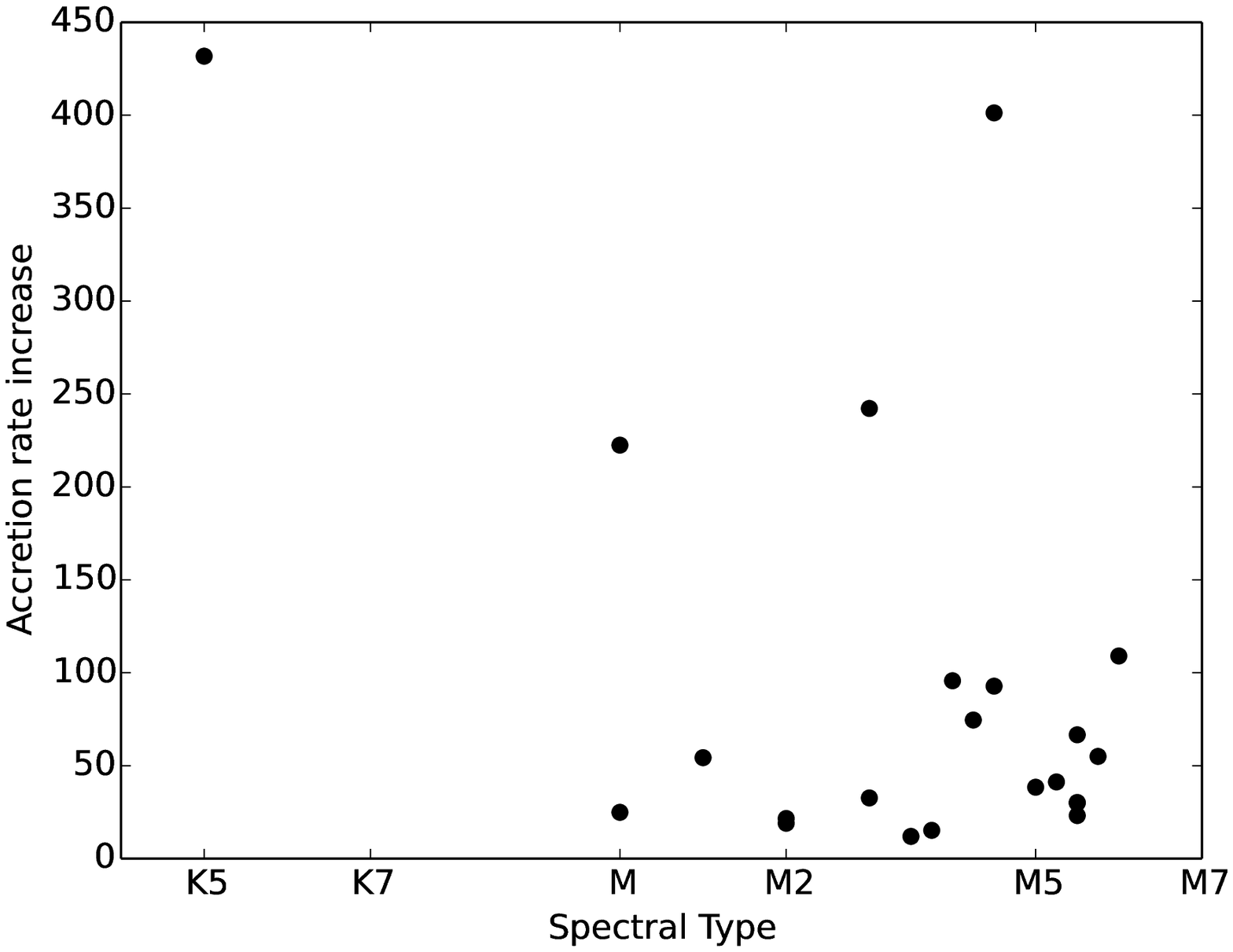}
\caption{Estimated increase in accretion rate during bursts, as a function of stellar spectral type.}
\label{accincrease}
\end{figure}

Given our estimates of $L_*/L_{\rm acc}$ and $r$ as measured from the $K2$ photometry, we have calculated the increase in mass accretion rate during bursts for each of our sources. Figure~\ref{accincrease} illustrates the resulting values as a function of spectral type, in cases where this is known to a subclass or better. We find that they range from $\sim$10 to over 400 for the most prominent burst events.

\section{Stellar properties of the accretion burst objects}

\subsection{Spatial distribution}

\begin{figure*}
 \epsscale{1.0}
 \plotone{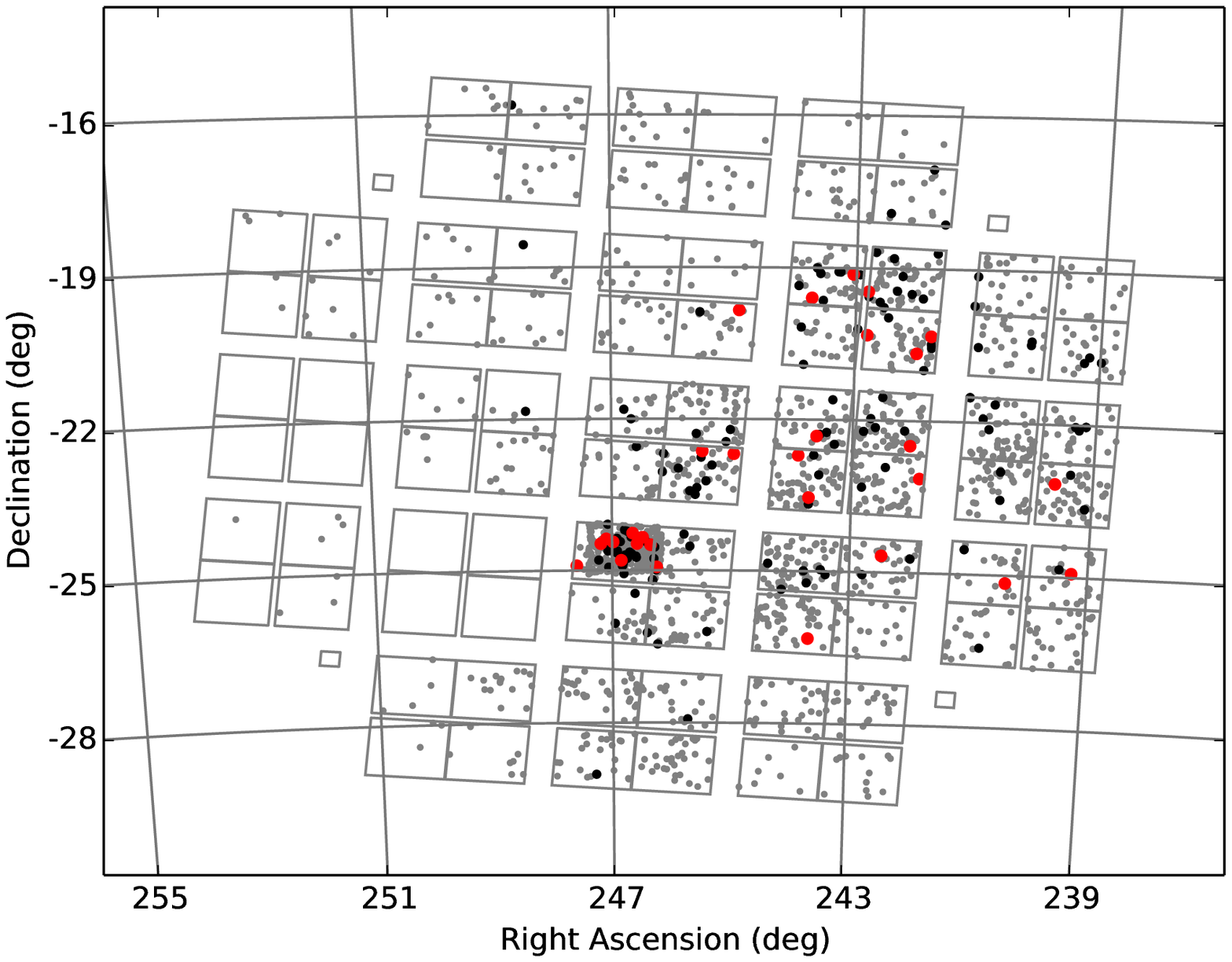}
 \plotone{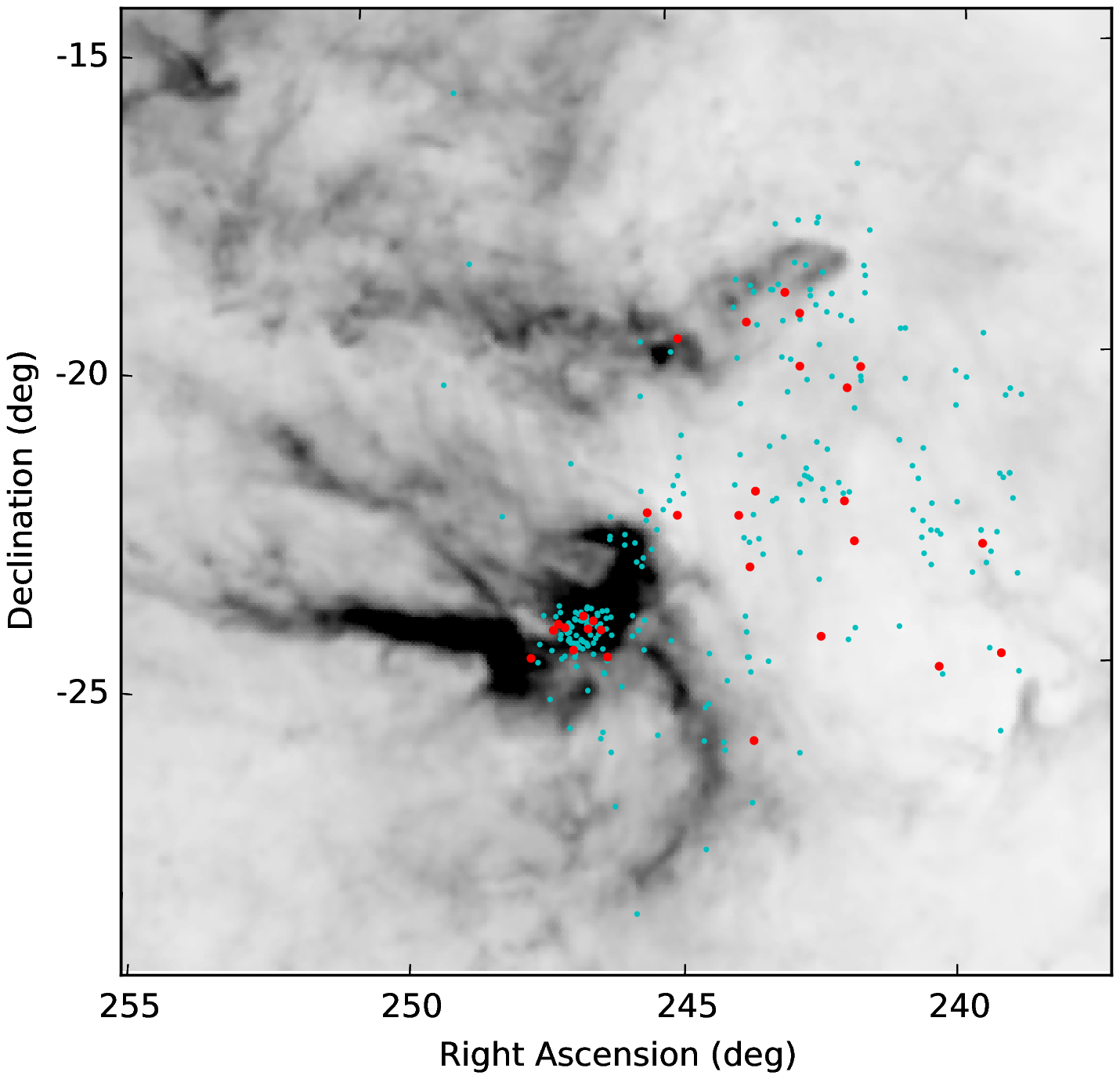}
 \caption{Top: Spatial distribution of young stars (gray), including variables (black) and specifically bursters (red), overlaid on the K2 field of view. The concentration of stars near RA=246.8, Dec=-24.6 is the $\rho$~Oph cluster. Bottom: Young stars with disks (cyan) and bursters (red) overlaid on the \citet{2011ApJ...737..103S} extinction map of the $\rho$~Oph region.}
 \label{radec}
 \end{figure*}

The sample we have analyzed here is likely a mixture of ages
from young ($<1-2$ Myr) but still optically visible stars associated with the young $\rho$ Oph cloud,
to somewhat older (5-10 Myr) stars in the off-cloud Upper Sco region which still retain 
accreting circumstellar disks. Bursters as identified here make up $\sim$9\% of the disk-bearing young star sample in the $K2$/C2 dataset. 
We initially hypothesized that they would preferentially appear in the compact, young $\rho$~Oph cluster, as opposed
to the more dispersed and older Upper~Sco region. But the spatial distribution shown in
Figure~\ref{radec} surprisingly reveals equal proportions of bursters in the two regions. Furthermore, there is no correlation with global extinction measures. This suggests that either the young population extends from $\rho$~Oph out into the surrounding areas, or that the burster
phenomenon is less dependent on age than on disk properties such as mass. \citet{2011AJ....142..140E} found evidence for an intermediate age population ($\sim$3~Myr) of YSOs outside of the main $\rho$~Oph cloud core, but within the main L~1688 cloud. Furthermore, 
\citet{2005AJ....130.1733W} found a negligible age difference between sets of Upper~Sco and L~1688 association members;
both regions were estimated to be $\sim$3~Myr old. 

 \subsection{Spectral types}
\begin{figure}
 \epsscale{1.30}
 \plotone{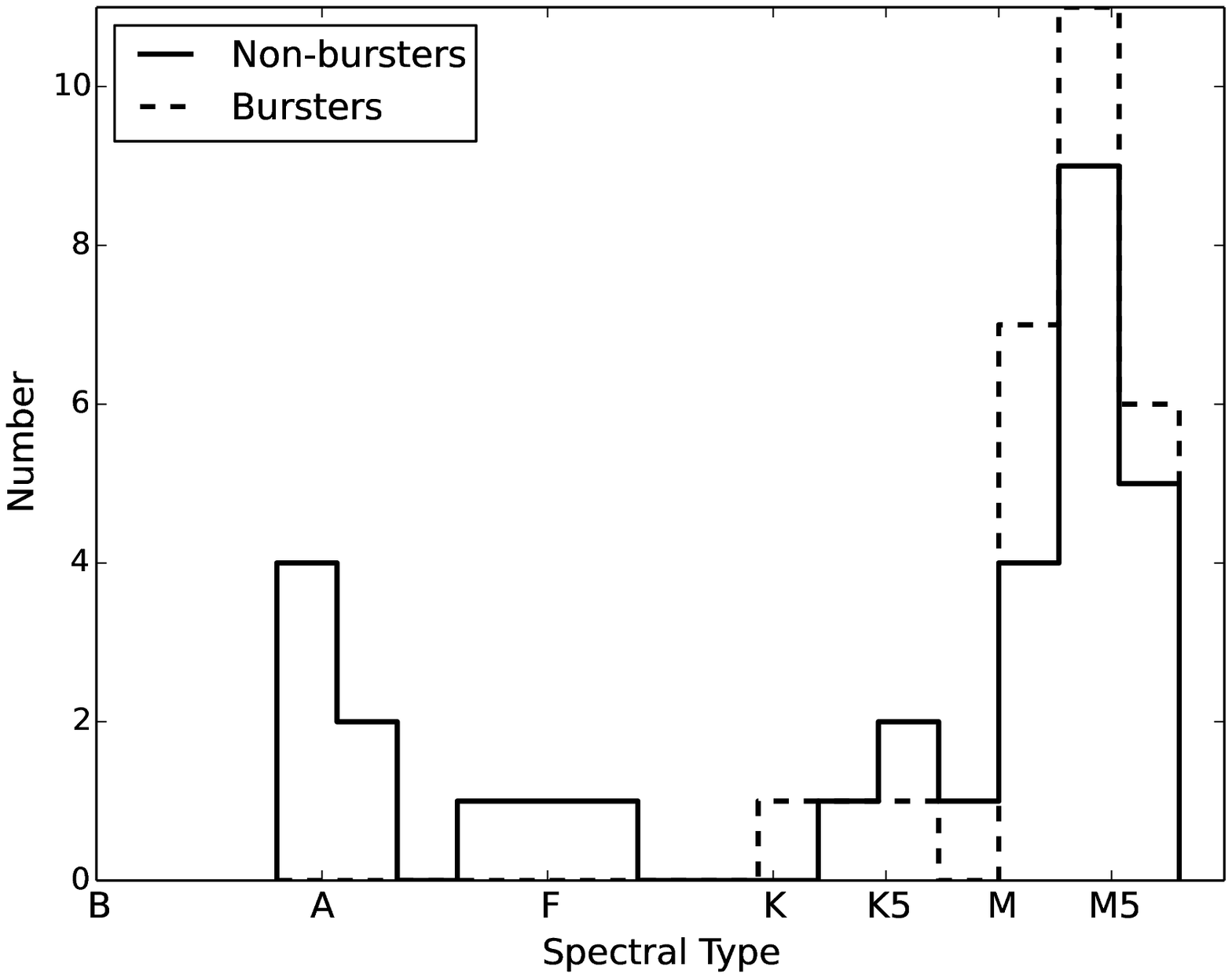}
 \caption{Spectral type distributions for bursting and non-bursting young disk-bearing stars observed in K2 Campaign 2. Two distributions are very different for the B--F range, but indistinguishable for K and M spectral types.}
 \label{spts}
 \end{figure}

 To assess the mass distribution of stars displaying bursts, we have gathered spectral types from the literature.
 Those found for the bursters are displayed in Table~1. We have also selected a control sample of non-bursting stars from \citet{2012ApJ...758...31L}. That work presents spectral types for hundreds of USco members; we cull the list to include only non-bursting, inner disk-bearing stars observed in K2 Campaign 2. The resulting set of 31 objects has spectral types ranging from B8 to M8, as shown in Figure~\ref{spts}. We compare this against the distribution of spectral types for the bursters identified here, finding a significant difference only for the B--F spectral types. There are early type stars in the sample but none of their variability is of the bursting type. K2 light curves for these objects show mainly low-level quasi-periodic modulation. However, the spectral type distributions of bursters and non-bursters is very similar for the K--M range. Statistically, the two sets are indistinguishable here. Thus we conclude that the preponderance of late-M spectral types among bursters is consistent with the increased fraction of young stars with disks at low masses. 
 
 \subsection{Multiplicity}
 The presence of close stellar companions to the young stars in our sample may influence their variability properties. Indeed it has been hypothesized that, for high-amplitude outbursting stars such as FUors, events could be triggered by the presence of a perturbing companion \citep{2004ApJ...608L..65R,2004MNRAS.353..841L}. However, support for this idea is mixed \citep{2016arXiv160503270G}. It is nevertheless worthwhile to check which, if any, of our burster sample may be in binary systems. We have vetted over half of the sample for multiplicity by examining the available high-resolution imaging as well as obtaining new speckle observations.
 
 \subsubsection{Literature assessment}
 
We mined the literature for adaptive optics and speckle imaging of these targets, and the details are tabulated as part of the individual object commentary in the Appendix. Only three objects have evidence for a close companion published in the literature: 
\begin{itemize}
\item EPIC 204906020 (2MASS J16070211-2019387), 
\item EPIC 203786695 (2MASS J16245974-2456008), 
\item EPIC 203905625 (2MASS J16284527-2428190).
\end{itemize} The first is a close binary \citep[55~mas or 8 AU;][]{2012ApJ...757..141K}. Likewise, the second has a companion at $\sim$15~AU (100.4~mas) \citep{2002AJ....124.1082K}.  The last of these has an even closer companion at $\sim$2~AU (13~mas) \citep{1995ApJ...443..625S}. Interestingly, all three systems also have wider ($>$50 AU) companions. None of their light curves shows any hint of a periodicity. One additional object, EPIC 204342099 (2MASS J16153456-2242421), may have a 274~AU separation companion \citep{1992AJ....103..549G}, but it is not clear whether the two stars are bound. EPIC 204830786 (2MASS J16075796-2040087) is associated with another star 21.5\arcsec\ away ($>$3000 AU separation) but has not been surveyed for closer companions. 

Other than these five objects, 9 other bursters in our sample have been surveyed for multiplicity, at a variety of sensitivities and separations: 
\begin{itemize}
\item EPIC~204226548 (2MASS~J15582981-2310077), 
\item EPIC~203899786 (2MASS~J16252434-2429442),
\item EPIC~203935537 (2MASS~J16255615-2420481),
\item EPIC~203905576 (2MASS~J16261886-2428196),
\item EPIC~203954898 (2MASS~J16263682-2415518),
\item EPIC~203822485 (2MASS~J16272297-2448071),
\item EPIC~203913804 (2MASS~J16275558-2426179),
\item EPIC~203928175 (2MASS~J16282333-2422405), 
\item EPIC~203794605 (2MASS~J16302339-2454161).
\end{itemize} No companions were found in these cases. However, the separations probed are very non-uniform and range from 10~mas (1.5~AU) in some cases to 1--30\arcsec\ in others ($>$150~AU); details are provided in the Appendix individual objects section.
 
\subsubsection{DSSI targets} 
In addition to data compiled from the literature, we also have speckle imaging observations of six bursters using DSSI (\S 2.3). For EPIC 204342099 (2MASS~J16153456-2242421), we recover the companion reported at 1.9\arcsec\ by \citet{1992AJ....103..549G}; however, we measure the separation to be 1.50\arcsec\ ($\sim$218~AU), and a magnitude difference of $\Delta$m=3.38 at the 880~nm band. This star previously had direct imaging and aperture masking by \citet{2008ApJ...686L.111K} that ruled out further objects down to 240~mas (35~AU). EPIC 204830786 (2MASS~J16075796-2040087) has a previously noted possible companion at thousands of AU separation, but the DSSI observations otherwise support the hypothesis that this is a single star. EPIC~204440603 (2MASS~J16142312-2219338) has no previous multiplicity information, but the 880~nm image suggests a possible companion at a separation of ~0.1\arcsec. However, the lack of a similar detection at 692~nm and the faintness of this star makes speckle reconstruction challenging; thus the existence of such a companion remains indeterminate.
EPIC~204360807 (2MASS~J16215741-2238180) has no multiplicity information in literature, but we find it to be a 0.48\arcsec\ separation binary (70~AU), with $\Delta$m=0.74 at 880~nm. EPIC~203935537 (2MASS~J16255615-2420481) has no reported evidence for multiplicity, and we do not detect companions down to 0.1\arcsec\ separation (15~AU) at 4--5 magnitudes of contrast in the 692 and 880~nm bands. Finally, EPIC~203928175 (2MASS~J16282333-2422405) was reported by \citet{2015ApJ...813...83C} to host no companions down to 20 mas (3~AU) at 1-3 magnitude contrast; our observations support the lack of binarity, with no detections outward of 0.1\arcsec\ at $\Delta m\sim$ 4.4 magnitudes.

 \section{Time domain behavior}

In order to appreciate the diversity of the bursting behavior among our sample of objects,
we must go beyond just the $M$ and $Q$ metrics discussed above.  We quantified
the peak-to peak amplitudes of the bursters by first cleaning and normalizing the light curves and then computing the maximum-minus-minimum values. There are several ways to quantify light curve timescales. 

\begin{figure}
 \epsscale{1.3}
 \plotone{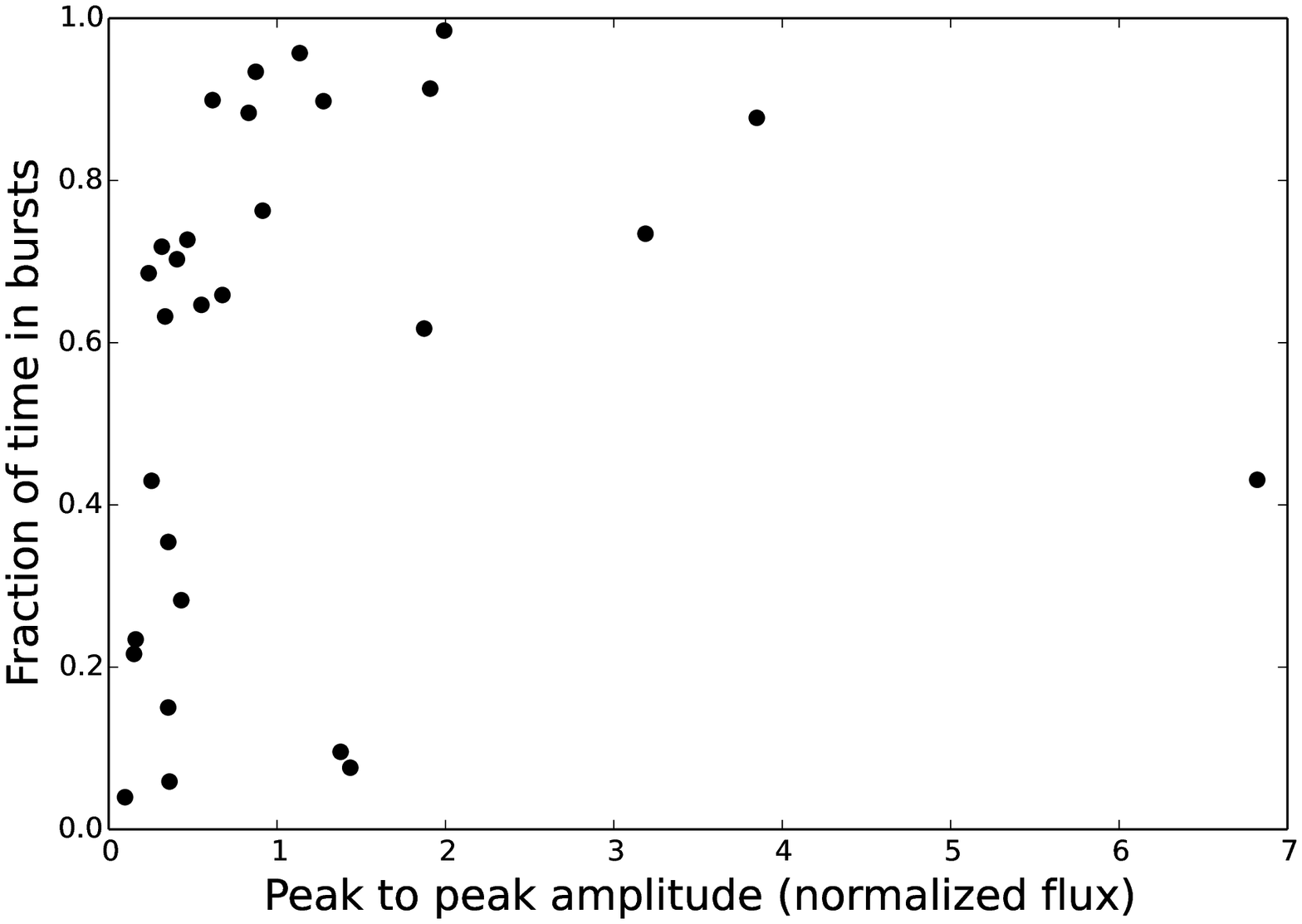}
 \caption{Maximum burst amplitude (in units of normalized flux) versus duty cycle, i.e., fraction of time spent bursting.}
 \label{ampburstfrac}
 \end{figure}

First, we measure the burst duty cycle, which is the fraction of time each object spends in a bursting state. This is by nature somewhat subjective, as bursts display a range of amplitudes and shapes. We identified bursting portions of each light curve by first fitting and removing a low-order median trend to the light curve. This flattens the ``continuum'' level from which bursts arrive.
We then measure the typical point-to-point scatter by shifting each point by one, subtracting from the original light curve, and dividing the standard deviation of the result by $\sqrt{2}$. Using this measure of scatter, we have found that burst behavior, as detected by eye, includes points that lie about 15 times the scatter above the minimum of the continuum-flattened light curve. We thereby selected bursting and non-bursting sections for each time series. This method only failed for three objects (EPIC 204397408/2MASS~J16081081-2229428, EPIC~205156547/2MASS~J16121242-1907191, and EPIC~203856109/2MASS~J16095198-2440197) that displayed intermittent quasi-periodic behavior that was picked up as bursts. We manually removed these light curve portions for the statistical analysis. 
In Figure~\ref{ampburstfrac} we display the peak-to-peak amplitudes versus duty cycle of each burster. The duty cycles exhibit a large range of values, from almost 100\% down to $\sim$10\%. Typically the light curves with the highest amplitudes have higher duty cycles of 60\% and above, with the exception of outlier EPIC 203954898/2MASS~J16263682-2415518. This object may represent a distinct form of bursting behavior. 

\begin{figure}
 \epsscale{1.3}
 \plotone{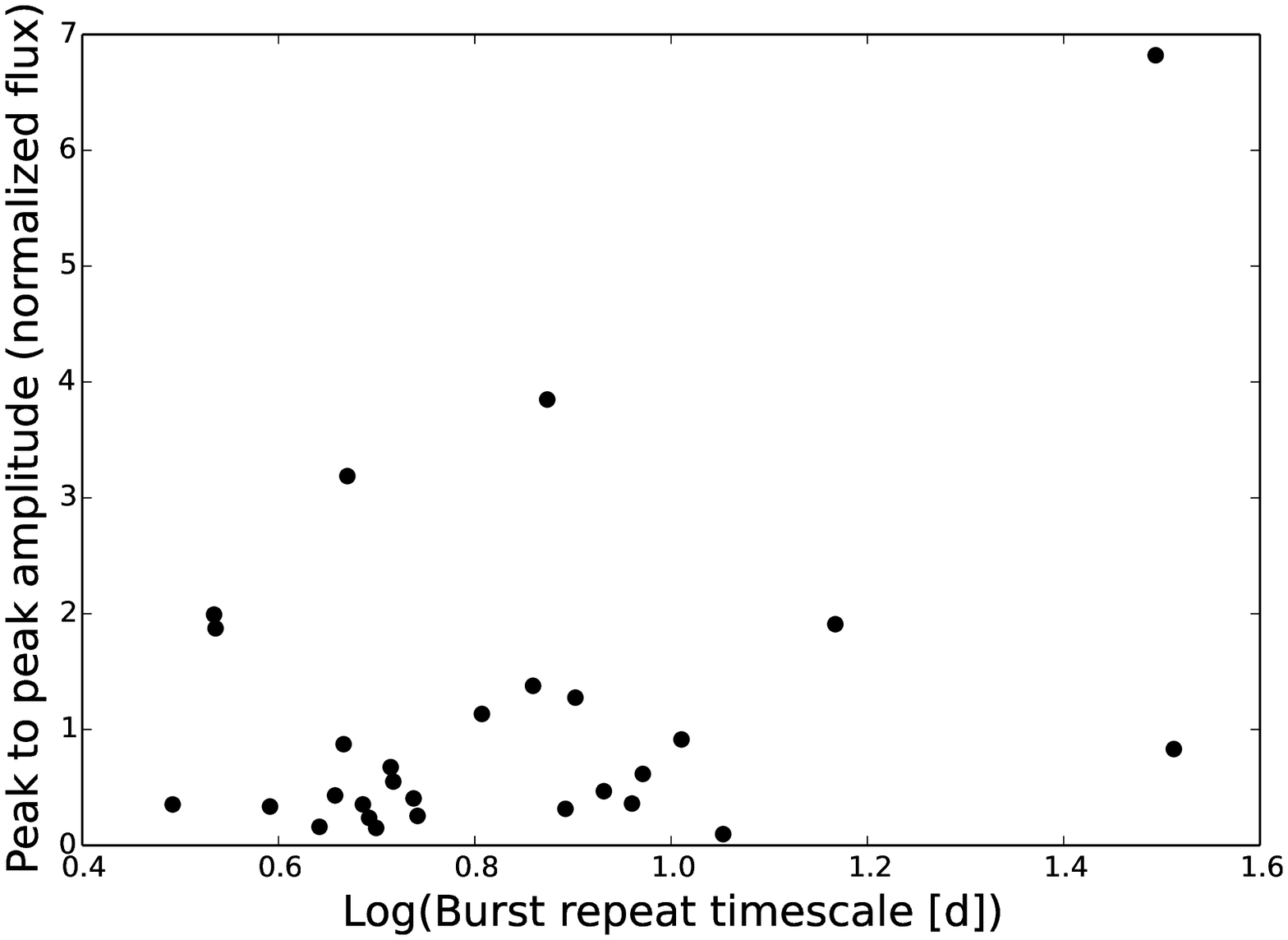}
 \caption{Burst repeat timescales versus peak-to-peak light curve amplitude, in normalized flux.}
 \label{timescaleamp}
 \end{figure}

We also quantify the burst timescales by applying a method similar to the one described in \citet{2014AJ....147...82C} (\S6.5 of that paper). In brief, this involves identifying peaks that rise above a particular amplitude level compared to the surrounding light curve. Once peaks are found, the median timescale separating them is computed. This procedure is repeated for a variety of amplitudes, from the noise level up to the maximum light curve extent. The result is a plot of timescale versus amplitude (e.g., Fig.~32 in \citet{2014AJ....147...82C}). Finally, we take as a ``representative'' timescale the value corresponding to an amplitude that is 40\% of the maximum peak-to-peak value (we note that this is different from the value of 70\% adopted in \citet{2014AJ....147...82C} and appears more appropriate for the burster light curves examined here). This computation only fails for the light curve of EPIC 203382255 (2MASS~J16144265-2619421), which displays only one burst event; here the timescale is indeterminate. In Figure~\ref{timescaleamp} we display the peak-to-peak amplitudes versus estimated timescale for each burster. Again, there is a large range of values, but no clear correlation with amplitude.

\begin{figure}
 \epsscale{1.3}
 \plotone{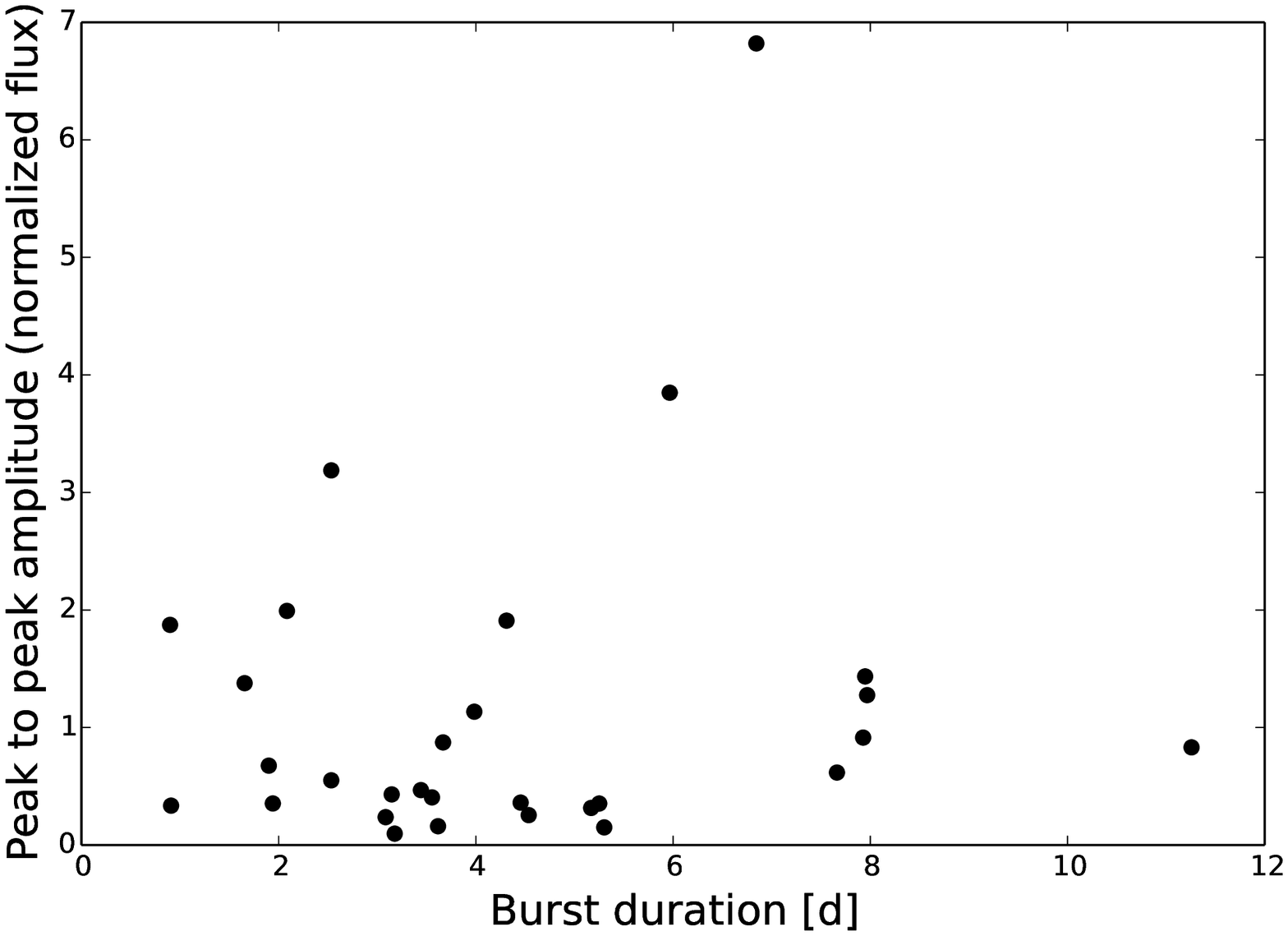}
 \caption{Burst durations versus peak-to-peak light curve amplitude, in normalized flux.}
 \label{durationamp}
 \end{figure}

\begin{figure}
 \epsscale{1.3}
 \plotone{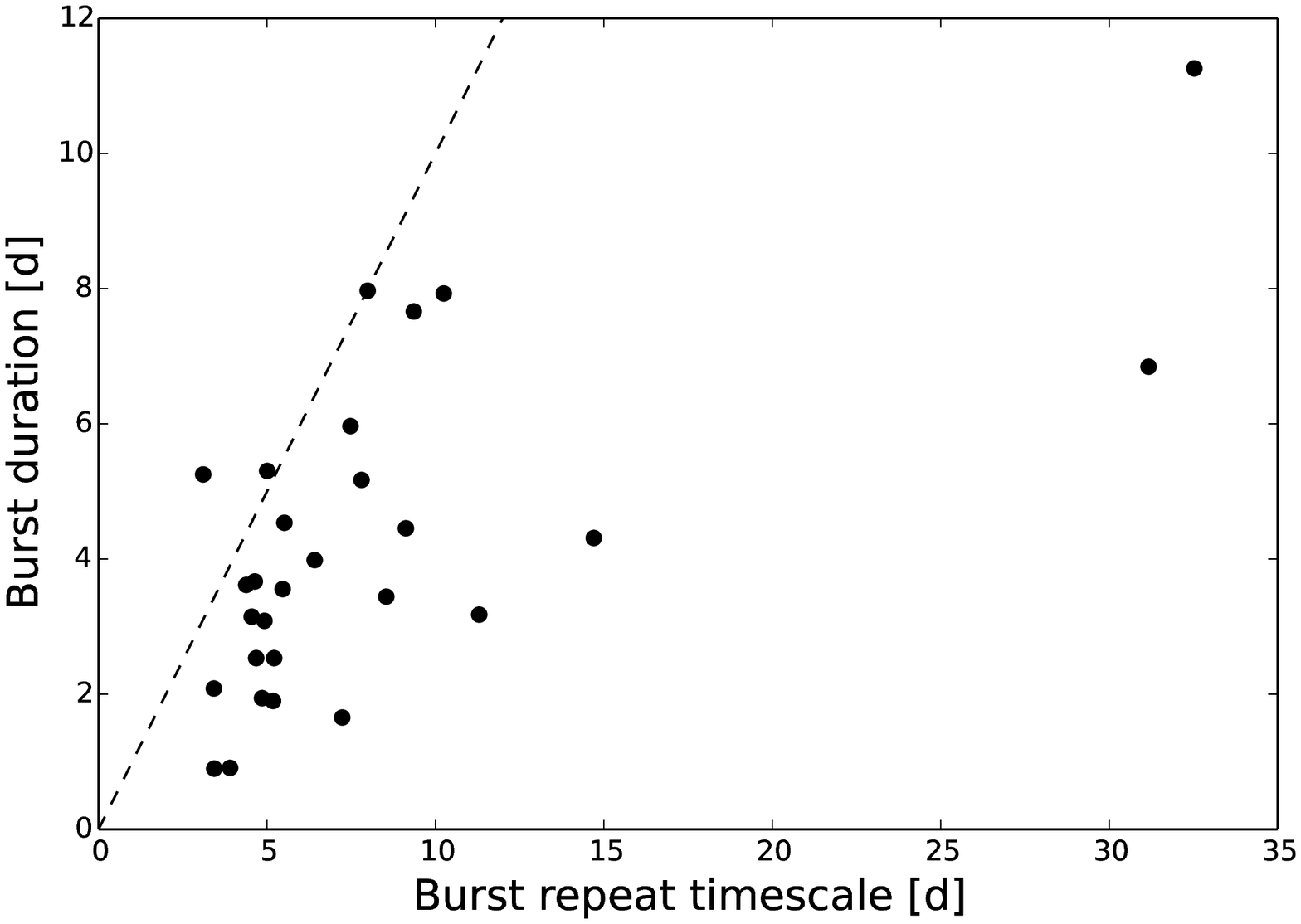}
 \caption{Burst repeat timescale versus durations. The dashed line indicates where these two quantities are equal. We have left out EPIC 203382255 (2MASS~J16144265-2619421) since it only has one burst and the repeat timescale is thus indeterminate.}
 \label{timescaleduration}
 \end{figure}

The burst duration is another way to quantify the observed events. This is a challenging measurement, as there is a superposition of bursts with varying widths and heights. We simplify as above by only considering peaks that rise to a level of at least 40\% of the maximum peak-to-peak value. For each peak, we identify the surrounding points that are more than 15 times the point to point scatter above the minimum light curve value (as was done for the burst duty cycle calculation). We then adopt the median burst duration of all such peaks in each light curve. The result is plotted against peak-to-peak amplitude in Figure~\ref{durationamp}. We also compare the durations with the repeat timescales in Figure~\ref{timescaleduration}. Here we find that burst duration is correlated with repeat timescale. This is somewhat expected since, by definition, a duration is typically larger than the average timescale between bursts. However, we observe a distinct lack of short bursts (duration $\leq$2 days) with long repeat timescales. This may be rooted in the physical mechanism of the bursts.

We have also generated periodograms to identify any repeating components in the light curves. Most bursters do not exhibit periodicity but instead adhere to stochastic behavior, with any quasi-periodicity quantified via the $Q$ statistic (see \S 3). Those that do show significant periodicities (as indicated by $Q\leq 0.61$ and/or a strong, isolated periodogram peak) are EPIC 203794605/2MASS~J16302339-2454161 ($P=4.5$d), EPIC 203899786/2MASS~16252434-2429442 ($P=6.0$d), EPIC 203928175/2MASS~J16282333-2422405 ($P=4.4$d), EPIC 203954898/2MASS~J16263682-2415518 ($P=20.8$d),  and EPIC 204347422/2MASS~J16195140-2241266 ($P=6.9$d). The measured periods are similar to the burst repeat timescales inferred above.
In addition, EPIC 203856109 (2MASS~J16095198-2440197), EPIC 204233955 (2MASS~J16072955-2308221), EPIC 204397408 (2MASS~J16081081-2229428), and EPIC 205156547 (2MASS~J16121242-1907191) display short-timescale ($P$ less than a few days) periodic behavior outside of their bursting states. These periods are more typical of the K2 $\rho$ Oph and Upper Sco sample as a whole \citep[see][]{Reb17} and likely measure stellar rotation, whereas those of the bursts are longer by factors at least several.

\begin{figure}
 \epsscale{1.3}
 \plotone{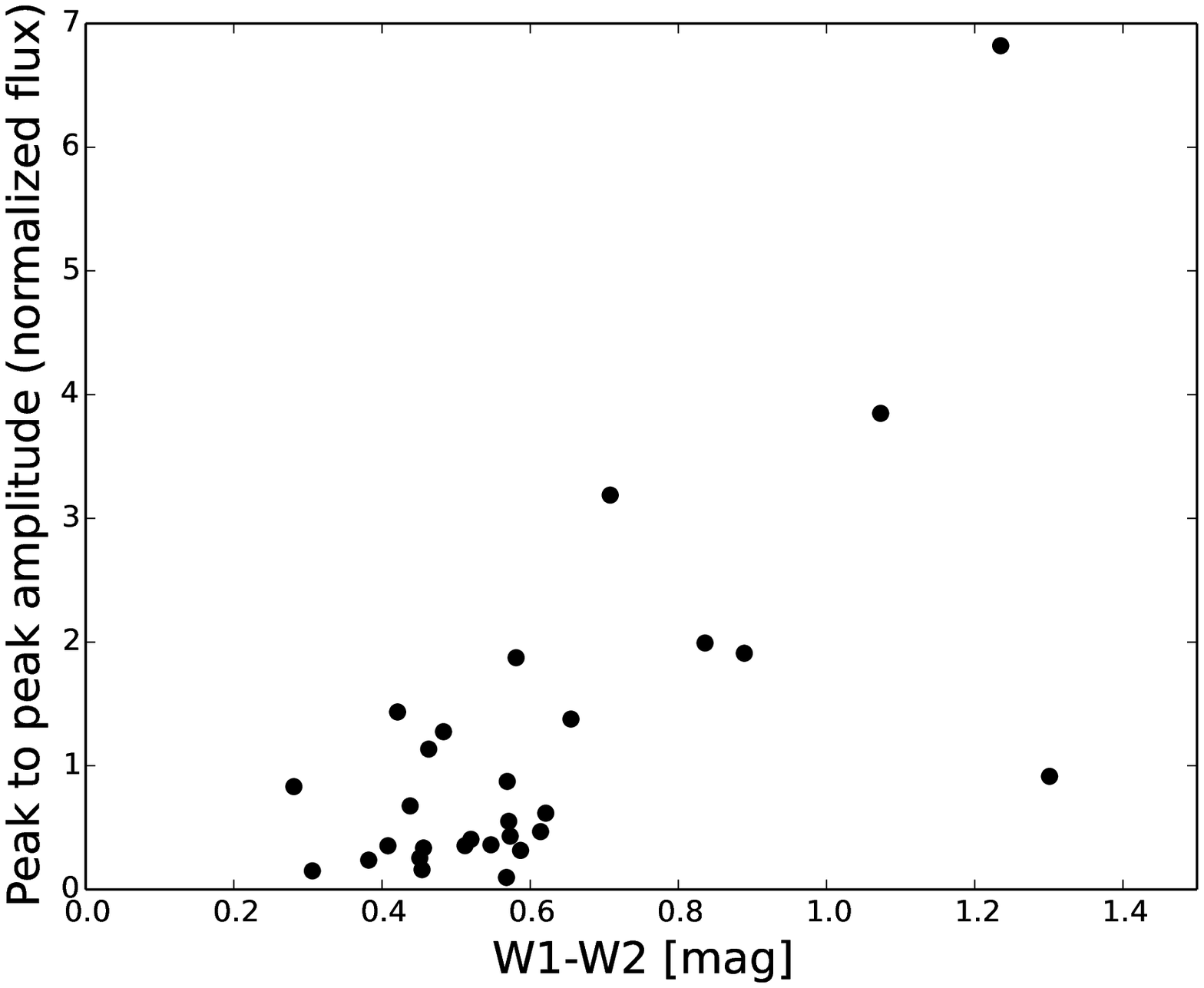}
 \caption{Peak-to-peak light curve amplitude for bursters is shown against the $W1-W2$ color. All included objects have $W1-W2$ values indicative of inner disk dust, but the highest amplitude bursters tend to have larger infrared excesses, with one prominent exception (EPIC 205008727/2MASS~J16193570-1950426; $W1-W2\sim 1.3$).}
 \label{W1W2amp}
 \end{figure}

Few of the timescale metrics show any relation to [circum]stellar properties, but one potential correlation stands out in 
peak-to-peak amplitude versus the infrared $W1-W2$ color (Figure~\ref{W1W2amp}), which is indicative of a dusty inner disk. These two quantities are correlated at a significance level of 4$\times 10^{-5}$ (Pearson $r$ coefficient of 0.69). This is also borne out in Figure~\ref{JKWISE}, which suggests that the dustiest objects have the highest light curve amplitudes.

\begin{deluxetable*}{llcccccccc}
\tabletypesize{\scriptsize}
\tablecolumns{8}
\tablewidth{0pt}
\tablecaption{Light curve metrics for bursting young stars in $K2$ Campaign 2}
\tablehead{
\colhead{EPIC id} & \colhead{2MASS id} & \colhead{Amplitude} & \colhead{Q} &  \colhead{M} &  \colhead{Timescale} & \colhead{Duration} & \colhead{Duty cycle} & \colhead{Period} \\ 
\colhead{} & \colhead{} & \colhead{(norm.\ flux)} & \colhead{} &  \colhead{} &  \colhead{[d]} & \colhead{[d]} & \colhead{} & \colhead{[d]} \\ 
}
\startdata
203382255 & J16144265-2619421 & 1.43 & 1.00 & -1.14 & $>$77.74 & 7.95 & 0.08 & - \\
203725791 & J16012902-2509069 & 0.68 & 0.87 & -0.29 & 5.18 & 1.90 & 0.66 & - \\
203786695 & J16245974-2456008 & 0.25 & 1.00 & -0.59 & 5.52 & 4.54 & 0.43 & - \\
203789507 & J15570490-2455227 & 0.36 & 0.97 & -0.41 & 9.12 & 4.45 & 0.06 & - \\
203794605 & J16302339-2454161 & 0.87 & 0.55 & -0.25 & 4.64 & 3.67 & 0.93 & 4.46 \\
203822485 & J16272297-2448071 & 0.62 & 0.84 & -0.29 & 9.36 & 7.66 & 0.90 & - \\
203856109 & J16095198-2440197 & 0.35 & 1.00 & -1.00 & 3.11 & 5.25 & 0.15 & - \\
203899786 & J16252434-2429442 & 1.13 & 0.61 & -0.83 & 6.42 & 3.98 & 0.96 & 5.95 \\
203905576 & J16261886-2428196 & 3.85 & 1.00 & -0.66 & 7.48 & 5.97 & 0.88 & - \\
203905625 & J16284527-2428190 & 0.32 & 1.0 & -0.31 & 7.80 & 5.17 & 0.72 & - \\
203913804 & J16275558-2426179 & 0.41 & 1.0 & -0.37 & 5.47 & 3.56 & 0.70 & - \\
203928175 & J16282333-2422405 & 3.19 & 0.54 & -0.66 & 4.68 & 2.53 & 0.73 & 4.39 \\
203935537 & J16255615-2420481 & 0.24 & 1.00 & -0.31 & 4.92 & 3.09 & 0.69 & - \\
203954898 & J16263682-2415518 & 6.82 & 0.61 & -1.35 & 31.17 & 6.84 & 0.43 & 20.83 \\
204130613 & J16145026-2332397 & 1.87 & 0.85 & -0.35 & 3.43 & 0.90 & 0.62 & - \\
204226548 & J15582981-2310077 & 0.47 & 1.00 & -0.53 & 8.54 & 3.44 & 0.73 & - \\
204233955 & J16072955-2308221 & 1.99 & 0.85 & -0.82 & 3.42 & 2.08 & 0.98 & - \\
204342099 & J16153456-2242421 & 0.83 & 0.91 & -0.72 & 35.53 & 11.26 & 0.88 & - \\
204347422 & J16195143-2241332 & 1.38 & 0.75 & -1.11 & 7.23 & 1.65 & 0.10 & 6.94 \\
204360807 & J16215741-2238180 & 0.55 & 0.87 & -0.49 & 5.21 & 2.53 & 0.65 & - \\
204397408 & J16081081-2229428 & 0.15 & 0.59 & -0.68 & 5.01 & 5.30 & 0.22 & 1.65 \\
204440603 & J16142312-2219338 & 0.43 & 1.00 & -0.93 & 4.55 & 3.15 & 0.28 & - \\
204830786 & J16075796-2040087 & 1.91 & 1.00 & -0.67 & 14.70 & 4.31 & 0.91 & - \\
204906020 & J16070211-2019387 & 0.34 & 0.93 & -0.47 & 3.90 & 0.91 & 0.63 & - \\
204908189 & J16111330-2019029 & 1.28 & 0.76 & -0.59 & 7.99 & 7.97 & 0.90 & 19.23 \\
205008727 & J16193570-1950426 & 0.91 & 1.00 & -0.68 & 10.25 & 7.93 & 0.76 & - \\
205061092 & J16145178-1935402 & 0.35 & 1.00 & -0.51 & 4.85 & 1.94 & 0.35 & - \\
205088645 & J16111237-1927374 & 0.10 & 1.00 & -0.53 & 11.30 & 3.18 & 0.04 & - \\
205156547 & J16121242-1907191 & 0.16 & 1.00 & -1.01 & 4.38 & 3.62 & 0.23 & - \\
\enddata
\tablecomments{\label{tab:statstable} We tabulate basic statistical properties of the burster light curves. $Q$ and $M$ are discussed in \S3 as well as Cody et al.\ (2014). Amplitudes represent peak-to-peak measurements.
}
\end{deluxetable*}

\section{Discussion and Summary}
$K2$ data from Campaign 2 have probed the optical burst properties of young stars
on time scales ranging from hours up to several months, with approximately 8--10\% ($\pm$2\%) of strong disk sources exhibiting burst behavior. This is roughly in agreement with the 13$^{+3}_{-2}$\% fraction found for NGC~2264 \citep{2014AJ....147...82C,2014AJ....147...83S}. It is possible that an even larger fraction of young stars undergo bursting, but are not detected as such if the amplitude is low (i.e., $<100$\% peak to peak). Burst behavior could in some cases be masked by other types of variability.

Burster stars host inner circumstellar disks, as evidenced by {\em WISE} excesses and SEDs (Figure~\ref{seds}). Furthermore, there is a positive correlation between the $W1-W2$ color and light curve peak-to-peak amplitude. Stronger inner disk excesses appear to be associated with bigger bursts. Most of these objects have exceptionally strong H$\alpha$ emission, as well as other Balmer lines, \ion{He}{1}, and \ion{Ca}{2} in emission, as is typical for strongly accreting stars. Viewing geometry may play a role in setting the observability of the bursting phenomenon, but thus far we only have disk inclination constraints from ALMA on two sources in the sample, both of which are tilted at $\sim$42$\pm$15\arcdeg.

Members of the bursting sample typically exhibit multiple discrete brightening events, some lasting up to or just over one week. Time domain properties of the bursters are diverse, with some stars exhibiting a nearly continuous series of bursts and others displaying one or two isolated episodes superimposed on otherwise lower amplitude or quasi-periodic behavior. The majority of these light curves show flux variations of less than a factor of two and are similar to the objects in the $\sim$3~Myr NGC~2264 highlighted by \citet{2014AJ....147...83S}. However, seven of 29 bursters display discrete brightenings of more than 100\% on day to week timescales, unlike most heretofore classified young stellar variables that we are aware of. Of particular interest is the subset of these for which the bursts repeat quasi-periodically: EPIC~204347422 (2MASS~J16195140-2241266), EPIC~203928175 (2MASS~J16282333-2422405), and EPIC~203954898 (2MASS~J16263682-2415518). It is unclear as to what physical phenomenon sets the periods of 6.9~d, 4.4~d, and 20.8~d (respectively) in these cases; it appears to be relatively independent of stellar mass, as indicated by spectral type. There also is a lack of evidence for binarity in most of the bursters, although in some cases limits from imaging are relatively shallow. We thus speculate that burst events are not triggered by any companion, but rather by a repetitive interaction between the stellar magnetic field and inner disk.

From the theoretical perspective, magnetically channeled accretion is not predicted as steady in numerical simulations. An inner disk is truncated at a balance point between the inward pressure from accretion and the outward pressure of the magnetosphere. This produces variable and possibly cyclic mass flow due to instabilities and pulsational behavior.  Such variations in the mass loading of accretion columns are modeled under different physical scenarios by, e.g., \citet{1995MNRAS.275..244L}, \citet{1999ApJ...524..159G}, \citet{2002ApJ...578..420R,2004ApJ...610..920R,2005ApJ...635L.165R}, and \citet{2010MNRAS.406.1208D,2012MNRAS.420..416D}. 

The disk corotation radius ($r_c$) and the magnetospheric radius ($r_m$) are critical
in determining the regime of accretion under which a star-disk system falls. When $r_m<r_c$, gas at the inner disk edge rotates faster than the star and its magnetosphere, causing it to flow along magnetic field lines onto the central star. As long as $r_m>0.7r_c$, ``stable'' accretion occurs \citep{2016MNRAS.459.2354B} and gas follows two funnel streams onto the star \citep{2003ApJ...595.1009R,2004ApJ...610..920R}. For smaller values of $r_m$, the Rayleigh-Taylor instability sets in and accretion becomes chaotic, with many tongues of matter extending from the inner disk to the stellar surface \citep{2008ApJ...673L.171R,2008MNRAS.386..673K}. Numerous hot spots are present on the stellar surface, and the associated light curves display irregular bursting. This regime may be responsible for the large subset of bursters that we observe with high duty cycles.

When $r_m$ is up to a factor of two larger than $r_c$, on the other hand, gas in the inner disk rotates slower than the magnetosphere and is accelerated azimuthally. Models \citep[e.g.,][]{2012MNRAS.420..416D} predict a ``trapped disk" regime in which relatively continuous accretion bursts occur on short timescales. 
When $r_m>>r_c$, material tends to be flung out azimuthally in what is known as the ``strong propeller" regime \citep{2003ApJ...588..400R}. The light curves of several of our objects (EPIC 203954898/2MASS~JJ16263682-2415518 and EPIC 204347422/2MASS~J16195140-2241266) strongly resemble the simulated mass flow variations predicted by \citet{2013A&A...550A..99Z} and \citet{2014MNRAS.441...86L} for propeller behavior. In this scenario, the matter accretes episodically in three phases: First, material accumulates at the inner disk boundary, unable to flow inward. The magnetosphere is compressed toward the stellar surface. Next, compression reaches the point at which the magnetosphere can no longer withstand the gravitational forces on the accumulated material; gas accretes rapidly onto the star in a funnel flow. Finally, the compression pressure is relieved, accretion ceases, and the magnetosphere is able to expand outward again. Following this sequence, the cycle repeats. The time between bursts is longer than in other regimes, and the accretion rate can follow a flare-like pattern (rapid rise; slower decline) as a function of time. Strong outflows may also be present. EPIC~203382255 (2MASS~16144265-2619421) shows outflow signatures and also has the longest repeat timescale, making it an additional propeller regime candidate. In contrast, a few other objects with outflow-related spectral lines do not display the expected alternating burst and quiescence pattern.

Further investigation of these targets is necessary to estimate magnetic and corotation radii and compare with the theoretical expectations. The ratio $r_m/r_c$ depends on mass accretion rate, magnetic field strength, and stellar spin period-- parameters that we are unable to determine independently for this sample of burster stars.  In the meantime, the burst timescales measured for our sample provide perhaps the best indication of the physical mechanisms at work. Duty cycles range from 10\% and all the way up to nearly 100\%-- suggesting that we are seeing different modes of accretion from continuous to episodic and less frequent.  The possible correlation between burst duration and repeat timescale (Figure~\ref{timescaleduration}) may imply a relationship between mass loading timescale and accretion rate. 

A major open question stemming from this work is what the origin of the day to (multi-)week repeat timescales that we observe is, and how it relates to young outbursting stars with much longer duty cycles (e.g., EXors and FUors). Prominent examples from the literature include V899 Mon, which has had repeated bursts separated by $\sim$1 year \citep{2015ApJ...815....4N}, and V1647 Ori, which repeats at $\sim$2 years \citep{2009AJ....138.1137A}. On even longer timescales, other EXor type stars tend to outburst once or twice per decade. Unlike the $K2$ objects, the longer among these time scales are thought to be related to the viscous time scale in the disk. Here, the material drains inward and must undergo replenishment before the next instability-driven episode can occur. It has been speculated that the frequency and amplitude of outbursts may be set by the accretion rate, with younger and higher amplitude outbursting objects accreting more rapidly \citep[$\dot{M}\sim 10^{-7}$ to a few $10^{-4}$~$M_\odot$~yr$^{-1}$;][]{2014prpl.conf..387A}. We would then expect accretion rates for our own sources to be somewhat lower. This does indeed seem to be the case, as the median accretion rate where available for bursters is 10$^{-8.1}$~$M_\odot$~yr$^{-1}$ (for exact values, see the notes on individual objects in the Appendix). We have found (Section~4.2) that this increases one to two orders of magnitude during the most extreme $K2$ light curve peaks. Further spectroscopic measurements during times of definitive bursting may confirm these estimates.

In summary, the $K2$ mission is providing an unprecedented view of the time domain properties of young stars, and showing that the optical photometric manifestations of accretion phenomena take on a wide variety of timescales and amplitudes. Follow-up observations, including spectroscopy, should be carried out to investigate changes in spectral emission and inner disk structure in the highest amplitude objects presented here.

\acknowledgements

The work of A.M.C was supported by a NASA/NPP fellowship. We thank the referee for useful feedback that improved this paper. We acknowledge Luisa Rebull for calling our attention to one of the burst type objects that we had overlooked in our initial examination. Thanks also to Nic Scott for assistance with DSSI observations on the Gemini South telescope.

This paper includes data collected by the $K2$ mission. Funding for the $K2$ mission is provided by the NASA Science Mission directorate. The spectroscopic data were obtained at the W.M. Keck Observatory, which is operated as a scientific partnership among the California Institute of Technology, the University of California and the National Aeronautics and Space Administration. The Observatory was made possible by the generous financial support of the W.M. Keck Foundation. These results are also based on observations obtained as part of the program GS-2016A-Q-64 at the Gemini Observatory, which is operated by the Association of Universities for Research in Astronomy, Inc., under a cooperative agreement with the NSF on behalf of the Gemini partnership: the National Science Foundation (United States), the National Research
Council (Canada), CONICYT (Chile), Ministerio de Ciencia, Tecnolog\'{i}a e Innovaci\'{o}n Productiva (Argentina), and Minist\'{e}rio da Ci\^{e}ncia, Tecnologia e Inova\c{c}\~{a}o (Brazil).

This paper has utilized the SIMBAD and Vizier services, through which data from USNO-B, APASS, 2MASS, and WISE were collected. 

\bibliographystyle{apj.bst}
\bibliography{burst.bib}

\appendix

\section{Notes on individual objects}

Not all of our sample stars appear in the literature, but for those that have been studied previously, we highlight the key results here. We also incorporate information from our new speckle and spectroscopic observations.

\subsection{EPIC 203382255 / 2MASS 16144265-2619421}
This source is cataloged by \citet{2013MNRAS.431.3222L} as an astrometric
and photometric member, but is not otherwise studied in the
literature.  Our spectra suggest a spectral type of M4-M5.5
depending on the spectral range (earlier at bluer wavelengths).
Lithium is present with $W_{Li} = 0.30 A$ along with the H$\alpha$
emission reported in Table~\ref{tab:obstable}, \ion{He}{1},
weak NaD, and \ion{Ca}{2} are seen in emission.  Broad [\ion{O}{1}] is also
present.  The system is a clear accretion/outflow source.

\subsection{EPIC 203725791 / 2MASS J16012902-2509069} 
This star was first proposed as an Upper Sco member by \citet{2000AJ....120..479A}, based on $RIZ$ photometry. 
It was only recently confirmed by \citet{2015MNRAS.448.2737R}, based on significant lithium absorption. No other data has been reported for this star.

Our HIRES spectrum indicates a spectral type of M2 with lithium absorption strength $W_{Li}=0.3$ \AA. 
Strong H$\alpha$ emission is seen, as indicated in Table~\ref{tab:obstable}, along with lorentz-broadened \ion{He}{1}. NaD, weak narrow \ion{Fe}{2} as well as \ion{Mg}{1} emission, two-component \ion{Ca}{2}, and \ion{O}{1} emission. Very weak [\ion{O}{1}] is also exhibited.

\subsection{EPIC 203786695 / 2MASS J16245974-2456008}
This star, WSB 18, was first noted as part of Wilking et al.\ (1987)'s H$\alpha$ emission survey of the the $\rho$ Ophiuchi complex. Lithium absorption was detected by \citet{2011AJ....142..140E}, confirming youth. It is a visual binary with a separation of 1.1\arcsec\ (138~AU) and a 0.49 flux ratio \citep{1993A&A...278...81R}. Therefore, both components contribute to our $K2$ light curve. According to \citet{1997A&A...321..220B}, the primary has a spectral type of M2, whereas the secondary is M2.5. Both show H$\alpha$ in emission, but the primary does not appear to have a disk \citep{2006ApJ...636..932M}, whereas the secondary does. Further, \citet{2002AJ....124.1082K} found that the primary itself is a double star with 0.1\arcsec\ separation.

No detections of these sources at $>70$~$\mu$m were made with {\em Herschel}. In the submillimeter, \citet{2007ApJ...671.1800A} put an upper limit of 0.003~$M_\odot$ on the dust mass surrounding this source, although it is unclear which components were included.

\subsection{EPIC 203794605 / 2MASS J16302339-2454161}
According to \citet{1987AJ.....94..106W}, this star displayed an H$\alpha$ emission intensity of 3 on a scale of 1 to 5 (where 1 is weak and 5 is strong). Simon et al.\ (1995) performed a direct imaging search for companions in the 0.005-10\arcsec\ separation range, but did not find anything. Neither were any companions detected in the Catalog of High Angular Resolution Measurements \citep[CHARM;][]{2002A&A...386..492R}, its successor CHARM2 \citep{2005A&A...431..773R}, or the \citet{2005A&A...437..611R} speckle imaging survey of $\rho$ Oph. For the latter, no companions were detected down to 0.04 (0.14) times the stellar brightness at 0.5\arcsec\ (0.15\arcsec) separation. This star was included in the {\em Spitzer} c2d legacy survey.

Our HIRES spectrum is veiled, but consistent with a spectral type of M3.5-M5. Lithium is present with strength $W_{Li} = 0.32 \AA$. H$\alpha$ emission as reported in Table~\ref{tab:obstable} is strong with several components and a blueward-displaced central absorption. Additional emission includes \ion{He}{1} with lorentzian wings,
broad NaD, many multi-component \ion{Fe}{2} lines, strong \ion{Ca}{2} with multiple component and \ion{O}{1} 8446.  Among the outflow lines, only multi-component [\ion{O}{1}] is seen.

\subsection{EPIC 203822485 / 2MASS J16272297-2448071} This star, WSB 49, is in the $\rho$~Ophiuchus cluster. It was first detected as part of Wilking et al.\ (1987)'s H$\alpha$ emission line survey. It is also an IRAS source \citep{1989ApJ...340..823W}. It was first discovered as an x-ray source with the {\em ROSAT} High Resolution Imager \citep{2000A&A...359..113G}, and also noted XMM Newton observations by \citet{2010A&A...519A..34P}. \citet{2005AJ....130.1733W} confirmed its youth via detection of lithium absorption. \citet{2011AJ....142..140E} estimate a mass of 0.17~$M_\odot$. The star was surveyed for multiplicity, but no companions were detected down to $>0.15\arcsec$ separation at flux ratios of 0.06.

An infrared excess is detected with {\em Spitzer} \citep{2009ApJS..181..321E}.  \citet{2007ApJ...671.1800A} observed it in the submillimeter, classifying the disk as class II, with $<0.005$~$M_\odot$ of material. In addition, the disk is detected at 70~$\mu$m with Herschel/PACS \citep{2015A&A...581A..30R}. It was listed as a long timescale near-infrared variable by \citet{2014ApJS..211....3P}, while \citet{2014AJ....148..122G} detected $\sim$20\% variations in the mid-infrared with {\em Spitzer}.

\subsection{EPIC 203856109 / 2MASS J16095198-2440197} 
There is no previous literature on this source. Our spectrum indicates a spectral type of M5-M5.5 with
lithium at $W_{Li} = 0.61 \AA$. Moderate H$\alpha$ emission, as indicated in Table 1, is present and has a multi-component profile. Weak, also multi-component, \ion{He}{1} is also present, but no other emission lines.

\subsection{EPIC 203899786 / 2MASS J16252434-2429442} This star in the $\rho$~Oph cluster is also known as V852~Oph and SR~22. Its emission line spectrum was reported as early as the mid-20th century \citep{1949ApJ...109...92S}, with prominent hydrogen, \ion{Ca}{2}, and \ion{Fe}{2} noted. \citet{1972ApJ...174..401H} and \citet{1988cels.book.....H} included it in their emission 
line star catalogs. It is a {\em ROSAT} x-ray source \citep{1995ApJ...439..752C}. Irregular variability was also reported early on, by \citet{1972ATsir.728....5S} and \citet{1975PZ.....20..161F}. 
 
\citet{1994ApJ...420..837A} label the object as class III based on its 2.2--10~$\mu$m slope, and despite its previous classification as a classical T Tauri star by \citet{1979ApJS...41..743C}. It was relabeled as class II by \citet{1996AJ....112.2184G}, based on their 1.1--2.4~$\mu$m survey. \citet{2007ApJ...671.1800A} confirmed the disk with submillimeter observations, detecting 0.002~$M_\odot$ of material. 
Similarly, \citet{2013ApJ...773..168M} report 2.05~$M_{\rm Jup}$ of material, based on SCUBA-2 850~$\mu$m observations. \citet{2015A&A...581A..30R} detected the disk at 70~$\mu$m with {\em Herschel}, and reported that there is a gap in the spectral energy distribution that suggests a hole. 

The source is evidently a single star.  \citet{2005A&A...437..611R} ruled out any companions down to a flux ratio of 0.07 at a separation of 0.15\arcsec, and \citet{1996AJ....112.2184G} did not detect any radial velocity variations indicative of a spectroscopic binary. \citet{2015ApJ...813...83C} reported no companions down to 10~mas and contrasts of several magnitudes.

\subsection{EPIC 203905576 / 2MASS J16261886-2428196}
This star-- better known as VSSG 1-- boasts a disk that was first detected by IRAS \citep{1991ApJS...75..611C}. The mid-infrared slope, $\alpha$, is -0.4, making it a class II disk according to \citet{1994ApJ...420..837A} (although Wilking et al.\ 1989 earlier classified it as class I). \citet{2010ApJS..188...75M} used {\em SPEX} to measure an $n_{2-25}$ spectral index of -1.26, confirming the class II categorization. They also reported a small 10~$\mu$m silicate feature. \citet{2010ApJ...723.1241A} have observed this source's disk with the {\em Submillimeter Array} at 0.87~mm and estimated a dust mass of 0.029~$M_\odot$, along with an accretion rate of $10^{-7}$~$M_\odot$~yr$^{-1}$. H$_2$O was detected in the {\em Spitzer} IRS spectrum obtained by \citet{2010ApJ...720..887P}, along with HCN, C$_2$H$_2$ and CO$_2$. Salyk et al.\ (2011) confirm these detections and estimate a disk mass of 0.029~$M_\odot$. They infer a disk inclination of 53\arcdeg. Submillimeter observations with ATCA \citep{2010A&A...521A..66R} resulted in an estimated disk outer radius of 100--300~AU and a much lower dust mass of $4.5\times 10^{-5}$--$1.9\times 10^{-4}$~$M_\odot$. \citet{2014ApJ...782...51A} looked for mid-infrared variability in this source and concluded that it is {\em not} an EXor candidate. \citet{2006A&A...452..245N} report a fairly high accretion rate of 10$^{-7.19}$~$M_\odot$~yr$^{-1}$, and a Pa$\beta$ equivalent width of 8.9\AA\ in emission. According to the work of \citet{2005A&A...437..611R}, no companions are visible down to a flux ratio of 0.04 at a separation of 0.15\arcsec.

Our spectrum exhibits strong H$\alpha$ as indicated in Table~\ref{tab:obstable} with a double-peaked broad profile, as well as strong and broad \ion{Ca}{2} triplet as well as \ion{O}{1} 8446 emission. There is very little in the way of absorption, presumably due to heavy veiling, and spectral types from K7 to mid-M are plausible.  For the same reason, aggravated by low signal-to-noise in this region of the spectrum, there is no lithium measurement.

\subsection{EPIC 203905625 / 2MASS J16284527-2428190}
EPIC 203905625, also known as V853 Oph and SR 13, is a late-type star in $\rho$ Ophiuchus with reported spectral types from M2 to M4 \citep{2005AJ....130.1733W}. Accretion signatures include H$\alpha$ as well as calcium in emission. \citet{1976ApJS...30..307R} first reported strong veiling in the star's spectrum. \citet{1992A&AS...92..481B} found that the H$\alpha$ emission is variable (30--48\AA). \citet{2006A&A...452..245N} estimated an accretion rate of $10^{-8.31}$~$M_\odot$~yr$^{-1}$, from near-infrared spectra. They detected Pa$\beta$ in emission, at an equivalent width of -1.7\AA.

X-rays from this object were first detected with {\em Einstein} \citep{1983ApJ...269..182M}. 
A disk around this star was also observed, with {\em IRAS} \citep{1991ApJS...75..611C}. \citet{1994ApJ...420..837A} used 1.3mm observations to detect the disk; they classify it as class II,
based on a 2.2--10~$\mu$m slope of -0.8. \citet{2010ApJ...720..887P} detect H$_2$O, OH, HCN, and C$_2$H$_2$  in a {\em Spitzer} IRS spectrum of this target.
\citet{2015A&A...581A..30R} report Herschel detections at 70 through 500~$\mu$m; they classify the system as transitional, based on 12--24~$\mu$m data. 

The star is a multiple system, with a companion first detected at 0.4\arcsec\ separation via speckle imaging \citep{1993AJ....106.2005G}. \citet{1997MNRAS.284..257A} and \citet{2005A&A...437..611R} confirmed the 0.4\arcsec\ and a 0.238 flux ratio. \citet{1995ApJ...443..625S} conducted an IR imaging survey that revealed a closer companion at separation 13~mas.
\citet{2006ApJ...636..932M} found that both the primary and its 0.4\arcsec\ companion have class II disks. \citet{2007ApJ...671.1800A} measured a total disk mass of 0.01~$M_\odot$ using submillimeter observations, while \citet{2013ApJ...773..168M} measured a similar 7.81 $M_J$ from SCUBA-2 850~$\mu$m observations. 

Variability in this object was initially reported by \citet{1976ApJS...30..307R}. \citet{1982PZP.....4..127S} found optical variations from 12.6 to 14.7. \citet{1994AJ....108.1906H} observed V magnitude fluctuations from 12.83 to 13.52, while \citet{2007A&A...461..183G} reported $V$ magnitudes between 12.61 and 13.87. Likewise, \citet{1998AJ....116.1419S} conducted photometric monitoring, reporting a $V$-band amplitude of 0.91 magnitudes. The light curves are too sparsely sampled to identify any morphological features, although they are classified as ``irregular." The broadband {\em K2} light curve exhibits 0.2 mag events, suggesting that amplitudes are higher at bluer wavelengths. This is confirmed by Herbst et al.\ (1994)'s estimate of d$U$/d$V$: 2.39.

\subsection{EPIC 203913804 / 2MASS J16275558-2426179} This target is also known as SR 10 and V2059 Oph.  It originally appeared in Herbig \& Rao's (1972) catalog of emission line stars. Both \citet{1972ATsir.728....5S}, and \citet{1977IBVS.1248....1K} list it among their variable star compilations; \citet{1982PZP.....4..127S} labeled the variations as ``irregular."

H$\alpha$ emission at 40\AA\ was reported early on by \citet{1980AJ.....85..438R}. \citet{1983A&AS...53..291A} also observed strong H$\alpha$, as well as \ion{He}{1} and \ion{Fe}{2} in emission. \citet{1987AJ.....94..106W} noted this star in their H$\alpha$ emission survey, and \citet{2005AJ....130.1733W} again measured H$\alpha$ in emission as well as lithium in absorption. \citet{2006A&A...452..245N} estimated the accretion rate to be 10$^{-7.95}$~$M_\odot$~yr$^{-1}$; they detected Pa$\beta$ emission at an equivalent width of -5.6\AA. \citet{2015MNRAS.450.3559N} report a slightly lower accretion rate of 10$^{-8.3}$~$M_\odot$~yr$^{-1}$. 

\citet{1995ApJ...443..625S} conducted an imaging survey for binary companions at project separations from 0.005\arcsec\ to 10\arcsec\ down to a $K$ magnitude of 11.1. Likewise, \citet{2005A&A...437..611R} searched for companions at separations of 0.15\arcsec\ and 0.50\arcsec\ but did not identify any down to flux ratios of 0.04 and 0.02, respectively. \citet{2002A&A...386..492R,2005A&A...431..773R}, \citet{2003ApJ...591.1064B}, and \citet{2015ApJ...813...83C} did not detect companions either, down to 10--20~mas separation at several magnitudes contrast. Thus this object appears to be a single star.

EPIC 203913804/2MASS J16275558-2426179 is a {\em ROSAT} x-ray source \citep{1995ApJ...439..752C}. The star is also an IRAS source \citep{1989ApJ...340..823W}. \citet{1994ApJ...420..837A} identified a class II circumstellar disk, and \citet{2009ApJS..184...18G} later confirmed with {\em Spitzer} data. \citet{2007ApJ...671.1800A} studied the system in the submillimeter and deduced an upper limit on the disk mass of 0.005~$M_\odot$, based on the 1.3~mm flux. Similarly, \citet{2013ApJ...773..168M} found an upper limit of 5.0284~$M_{\rm Jup}$. \citet{2015A&A...581A..30R} detected the disk at 70~$\mu$m with Herschel. 

\subsection{EPIC 203928175 / 2MASS J16282333-2422405}
EPIC 203928175 is a star in the $\rho$~Ophiuchus region that is also known as SR~20~W \citep{1949ApJ...109...92S} and ROXC~J162823.4. \citet{2005AJ....130.1733W} report a spectral type of K5, along with variable H$\alpha$ emission and an equivalent width of 35\AA.

It was surveyed for binarity by \citet{2005A&A...437..611R}, but no companions were detected down to 0.08 times the stellar flux at a separation of 0.15\arcsec, or 0.04 times the stellar flux at 0.50\arcsec. \citet{2015ApJ...813...83C} also did not find any companions down to 20~mas separation, at contrasts of 1--3 magnitudes. With DSSI, we do not make any detections outward of 0.1\arcsec\ at $\Delta m\sim$ 4.4 magnitudes (flux contrast $\sim$0.02) in the 692 or 880~nm bands.

The object is encircled by a class~II disk, as reported by \citet{2009ApJS..181..321E}. The disk was subsequently detected at 70, 160, 250, 350, and 500~$\mu$m with Herschel by \citet{2015A&A...581A..30R}, who tentatively classified it as transitional. 

\subsection{EPIC 203935537 / 2MASS J16255615-2420481} This star is also known as SR~4 and V2058~Oph. It has a long history of photometric and spectroscopic study, dating back to Struve \& Rudkj{\o}bing's (1949) publication of emission line stars. \citet{1972ApJ...174..401H}, \citet{1987AJ.....94..106W}, and \citet{1988cels.book.....H} listed it in their catalogs of emission line stars. It has had a range of H$\alpha$ equivalent widths measured from 84\AA\ to 220\AA\, as well as a low vsin$i$ of $\sim$9~km~s$^{-1}$ \citep{1990AJ.....99..946B}. \citet{1993AJ....106.2024V} acquired blue spectra of this target, which revealed H($\beta$, $\gamma$, $\delta$) and Ca in emission, as well as significant veiling at 4450\AA. \citet{2007ApJ...669.1072E} measured the veiling factor, $r$, to be $\sim$1.5. \citet{2005AJ....130.1733W} detected lithium absorption and H$\alpha$ emission in this source. \citet{1996A&AS..120..229R} classified the H$\alpha$ emission line profile as type IIR, in which there are two peaks of similar height. The accretion rate is estimated by \citet{2006A&A...452..245N} to be a fairly high 10$^{-6.74}$~$M_\odot$~yr$^{-1}$, and the same  authors detected Pa$\beta$ in emission at an equivalent width of -19.0\AA. As suggested by \citet{2004AJ....127..420P}, the star may be the driver for a nearby Herbig Haro flow (HH 312) in the region.

The star is a {\em ROSAT} x-ray source. It is also an IRAS point source \citep{1989AJ.....97.1074I,1990ApJS...74..575W,1991ApJS...75..611C}. \citet{2001A&A...372..173B} also observed the class~II disk with ISO. \citet{2007ApJ...671.1800A} detected the disk at 850~$\mu$m with SCUBA, and they estimated a mass of 0.004~$M_\odot$ based on SED fitting. \citet{2013ApJ...773..168M} used SCUBA-2 to measure a larger disk mass of 9.4~$M_{\rm Jup}$ ($\sim$0.009~$M_\odot$).  {\em Spitzer}/IRS observations revealed a 10~$\mu$m silicate feature, with a typical equivalent width of 2.29~$\mu$m (Furlan et al.\ 2009). Interferometric data and modeling led to an inferred inner disk radius of 0.112~AU \citep{2007ApJ...669.1072E}. \citet{2008ApJ...673L..63P} estimated a very similar inner ring radius of 0.118~AU from near-infrared interferometry. \citet{2010ApJ...723.1241A} observed the disk with the Submillimeter Array and found a centrally peaked morphology. Their modeling predicts an inner disk radius of 0.07~AU and an inclination of 39\arcdeg; they infer an accretion rate of 10$^{-6.8}$~$M_\odot$~yr$^{-1}$, consistent with previous values. Millimeter and submillimeter ATCA observations by \citet{2010A&A...521A..66R} led to an outer disk radius of 100--300~AU and a dust mass of $\sim$2$\times$10$^{-5}$~$M_\odot$. The disk is also detected with {\em Herschel} at 160 and 250~$\mu$m \citep{2015A&A...581A..30R}. 

The object is a known variable, as originally reported by \citet{1972ATsir.728....5S,1982PZP.....4..127S} and \citet{1977IBVS.1248....1K}. It was followed up by \citet{1994AJ....108.1906H}, who found variations in the $V$ band of 12.73--12.93 during over 4000 days of monitoring. The amplitude was larger at blue wavelengths, with a typical d$U$/d$V$ of 2.4 magnitudes. \citet{1998AJ....116.1419S} reported a $V$-band amplitude of 0.41 magnitudes over both short (hour--day) and long (years) timescales, with no detectable periodicity. \citet{2007A&A...461..183G} monitored the star for over 7 years in the optical, finding a similar $V$-magnitude range of 12.60--13.09.

EPIC~203935537/2MASS~J16255615-2420481 is, to the best of our knowledge, a single star. Ghez et al.\ (1993)'s speckle imaging campaign did not reveal any companions down to 0.1\arcsec\ (0.2\arcsec), at a flux ratio of 17 (18). Neither did Simon et al.\ (1995)'s imaging survey, which was sensitive to separations of 0.005\arcsec--1\arcsec. Our own DSSI observations did not show any companions down to 0.1\arcsec\ separation at 4--5 magnitudes of contrast (flux ratio $\sim$0.02) in the 692 and 880~nm bands. \citet{2005A&A...437..611R} searched for companions in high resolution imaging but did not find any down to a flux ratio of 0.05 at a separation of 0.15\arcsec. Neither \citet{2003A&A...410..269M} nor \citet{2007A&A...467.1147G} detected any radial velocity variations indicative of spectroscopic binary status. Further high resolution imaging \citep{2002A&A...386..492R,2005A&A...431..773R} failed to reveal companions. 

\subsection{EPIC 203954898 / 2MASS J16263682-2415518}
This object was the subject of a multiplicity survey by \citet{2005A&A...437..611R}, but no companions were detected down to 0.05 (0.12) times the stellar brightness at 0.5\arcsec\ (0.15\arcsec) separation. Likewise, \citet{2007A&A...476..229D} did not detect any companion, and reported that two faint stars observed at 6.1\arcsec\ and 6.3\arcsec\ separation \citep{2004A&A...427..651D} are likely background objects. 

The star is a known X-ray emitter \citep{2000A&A...359..113G,2003PASJ...55..653I,2004ApJ...613..393G} and has a significant infrared excess, as indicated by ISO observations \citep{2001A&A...372..173B}, {\em Spitzer} data \citep{2009ApJS..181..321E,2009ApJ...696L..84C} and the AllWISE catalog \citep{2013wise.rept....1C}. The spectral index, $\alpha$, is 0.08, indicating a flat disk \citep{2009ApJS..181..321E}. 

A spectral type range of K7-M1 was reported by \citet{2011AJ....142..140E}. Our own spectrum from the Palomar 200-inch telescope Double Spectrograph suggests K8. We adopt M0. \citet{2005AJ....130.1145D} report an effective temperature of 3700$\pm$56~K, consistent with this spectral type. They also find a $v$sin$i$ of 27$\pm$4.7~km~s$^{-1}$. The mass has been estimated to be 0.18~$M_\odot$ \citep{2006A&A...452..245N}, which is very low considering the temperature and youth of the object. We suspect that several groups have confused the source with a neighboring star, 2MASS J16263713-2415599, which is $\sim$9\arcsec\ away. For example, \citet{2005AJ....130.1733W} list the object name ROXRA22, a ROSAT X-ray source at RA=16:26:36.9, Dec=-24:15:53 \citep{2000A&A...359..113G}. However, the \citet{2005AJ....130.1733W} coordinates match the companion star at RA=16:26:37.1, Dec=-24:15:59.9, and the listed spectral type is later, at M5. The companion paper by \citet{2011AJ....142..140E} provides the same effective temperature, luminosity, and mass estimate, but an earlier
spectral type range and lower extinction ($A_V$=4.9). Because of these discrepancies, we derive our own spectral data, apart from the $v$sin$i$ measurement. With the clear disk signatures, we can be confident that it is a young member of $\rho$~Ophiuchus. \citet{2006A&A...452..245N} list an accretion rate of $<10^{-9.71}$~$M_\odot$~yr$^{-1}$, based on near-infrared spectroscopy.

EPIC~203954898/2MASS~J16263682-2415518 was monitored in the mid-infrared with {\em Spitzer} as part of the Young Stellar Object Variability project \citep[YSOVAR;][]{2014AJ....148...92R}. \citet{2014AJ....148..122G} obtained 81 datapoints spread over 34 days at 3.6~$\mu$m. The mean magnitude in this band was 8.15, in line with previous brightness estimates \citep{2009ApJ...696L..84C,2009ApJS..181..321E}, and the standard deviation was 0.09 magnitudes. The star showed variability at the $\sim$0.05-magnitude level for the first 15 days of monitoring, followed by a $\sim$0.2-magnitude increase followed by a similar decrease over the remaining 20 days of observation. WISE data also display a $\sim$0.3-magnitude drop in the W1 band from February to August 2010. 

\subsection{EPIC 204130613 / 2MASS J16145026-2332397} This star, BV Sco, was listed as an irregular variable by \citet{1982PZP.....4..127S}. It has been erroneously classified as an RR Lyrae star in SIMBAD; this type of variability is inconsistent with the stochasticity seen in our $K2$ light curve. Like EPIC~204233955/2MASS~J16072955-2308221, it was labeled by \citet{2007MNRAS.374..372L} as a photometric non-member of Upper Scorpius (based on $ZYJHK$ data), before it was re-classified as a strongly accreting member \citep{2011A&A...527A..24L}. It showed both H$\alpha$ and \ion{He}{1} in emission (-108\AA\ and -3.0\AA\ equivalent widths, respectively). Furthermore, they detected the calcium triplet lines and forbidden \ion{O}{1} emission, suggesting outflows.

\subsection{EPIC 204226548 / 2MASS J15582981-2310077} 
Also known as USco CTIO 33, \citet{2000AJ....120..479A} first identified this star as a candidate Upper Sco member.  \citet{2002AJ....124..404P} confirmed its youth via measurement of lithium absorption; they also detected strong H$\alpha$ emission. With a spectral type of M3, it is estimated to be 0.36~$M_\odot$ by \citet{2007ApJ...662..413K}. These authors also searched for spectroscopic and wide (1--30\arcsec) binary companions, but did not find any. This star was detected as a {\em ROSAT} x-ray source (RX J155829.5-231026) by \citet{1998A&A...332..825S}. \citet{2006ApJ...651L..49C} were the first to detect an infrared excess, at 8 and 16~$\mu$m. Cieza et al.\ (2008) labeled it a transition disk source, with a mass of $<1.5\times 10^{-3}$~$M_\odot$ worth of material based on submillimeter data. No millimeter flux was detected by \citet{2012ApJ...745...23M}. \citet{2014ApJ...787...42C} report a submillimeter detection with ALMA, but the disk is unresolved. They constrain the dust mass to be $0.58\pm0.13$~$M_\oplus$. \citet{2009AJ....137.4024D} measure significant H$\alpha$ emission from a broad, flat-topped peak; their estimated accretion rate is $10^{-9.91}$~$M_\odot$~yr$^{-1}$. Similar values were derived by \citet{2010AJ....140.1444D}. Molecular gas is detected with Herschel/PACS (both C$_2$H$_2$ 
and HCN) by \citet{2013ApJ...779..178P}. \citet{2013A&A...558A..66M} estimate less than 0.9~$M_{Jup}$ worth of gas mass.

\subsection{EPIC 204233955 / 2MASS J16072955-2308221}

EPIC~204233955 is a spectral type M3 low-mass star in the Upper Scorpius association \citep{2011A&A...527A..24L}. It was initially classified as a photometric non-member by \citet{2007MNRAS.374..372L}, but \citet{2011A&A...527A..24L} found it to be an accreting source with strong emission lines, including H$\alpha$, \ion{He}{1}, and \ion{O}{1} forbidden lines. H$\alpha$ and \ion{He}{1} equivalent widths are -150\AA\ and -3.0\AA, respectively.
Mid-infrared data from the AllWISE survey \citep{2013wise.rept....1C} reveal a significant infrared excess in all bands, indicative of a disk. Other than the work of \citet{2007MNRAS.374..372L,2011A&A...527A..24L}, this star has not been studied in detail.

\subsection{EPIC 204342099 / 2MASS J16153456-2242421}
EPIC 204342099, otherwise known as VV~Sco, is a T Tauri star in the Upper Scorpius association. This object is also a known X-ray emitter from ROSAT observations \citep[1RXS J161534.0-224218;][]{2009ApJS..184..138H}, XMM Newton observations \citep[2XMM J161534.5-224241;][]{2012ApJ...756...27L}. \citet{1998A&A...333..619P} obtained low and medium resolution 
spectra of this star as part of an X-ray selected sample of candidate Upper Scorpius members. They reported a spectral type of M1 and confirmed its youth via lithium absorption.

EPIC~204342099 is a known disk-bearing source, originally discovered with IRAS \citep{1989AJ.....97.1074I}. It was studied in detail with {\em Spitzer}/IRS by \citet{2009ApJ...703.1964F}, who list it under the id 16126-2235. They find a very strong 10~$\mu$m silicate feature.

This star was first presented as a visual binary by \citet{1992AJ....103..549G}; the separation is 1.9\arcsec. 
\citet{2007ApJ...662..413K} measured a spectral type of M3.5 for the companion. It is not clear whether this object is a co-moving Upper Scorpius member or a serendipitous field object; if the former, then the separation is 274~AU.
With DSSI speckle imaging, we measure the separation to be smaller at 1.50\arcsec, with a magnitude difference of $\Delta$m=3.38 at the 880~nm band. \citet{2008ApJ...686L.111K} also searched for closer companions with direct imaging and aperture masking, but did not find any within 240~mas, at a magnitude difference of 2.8. 

Sparse variability data is available for EPIC~204342099/2MASS~J16153456-2242421. The star was first noted as variable by \citet{1975PZP.....2..221P}. \citet{1998A&AS..128..561B} monitored it for optical rotation signatures but did not detect any periodicities in the light curve.

Our HIRES spectrum suggests a spectral type of K9-M0 with lithium absorption present at strength $W_{Li}=0.45$ \AA. 
The H$\alpha$ emission is consistent with accretion (Table~\ref{tab:obstable}) and the profile exhibits a blueward
asymmetry along with a redshifted absorption notch against the emission.
Very weak and narrow profiles in \ion{He}{1}, \ion{Fe}{2}, and \ion{Ca}{2} are also present in our data, along with 
very weak and narrow [\ion{O}{1}].

\subsection{EPIC 204360807 / 2MASS J16215741-2238180} 
There is no previous literature on this source. As with other objects under consideration here, there is significant
veiling present with the spectral type changing from M2 in the bluer orders of our HIRES to possibly as late as M6 by about 8800 \AA. Lithium has strength $W_{Li} = 0.27 A$. H$\alpha$ emission as reported in Table~1 is very strong and there is  \ion{He}{1}, broad NaD, and many \ion{Fe}{2} lines. The \ion{Ca}{2} triplet has multiple components and  
\ion{O}{1} 8446 is present.  Of the outflow lines only weak [\ion{O}{1}] is seen.

With our DSSI speckle observations, we identify a companion at 0.48\arcsec\ separation with $\Delta$m=0.74 at 880~nm.

\subsection{EPIC 204397408 / 2MASS J16081081-2229428}
This object was first identified as a candidate USco member based on proper motion by \citet{2007MNRAS.374..372L} with \citet{2009A&A...504..981B} assigning it a membership probability of 99.9\% based on the USNO-B proper motion.
\citet{2008ApJ...688..377S} confirmed youth via variability and spectroscopy, assigning a spectral type of M5. They noted the star as an active accretor, based on a strong H$\alpha$ emission line. Likewise, \citet{2011A&A...527A..24L} also spectroscopically confirmed this object as a USco member. \citet{2012ApJ...745...56D} measured a radial velocity of about -11 km/s, slightly lower than the cluster mean, and a rotation velocity of $v\sin i=16.52\pm 4.05$.

\citet{2012MNRAS.420.2497R} analyzed {\em WISE} photometry for EPIC~204397408/2MASS~J16081081-2229428, concluding that it harbors a class II disk. \citet{2012ApJ...758...31L} also labeled it as a full disk.

\subsection{EPIC 204440603 / 2MASS J16142312-2219338} This very low mass star was classified by \citet{2007MNRAS.374..372L} 
as a photometric and proper motion member of Upper Scorpius based on UKIDSS data. Following up with the Anglo-Australian Telecope AAOmega spectrograph, \citet{2011A&A...527A..24L} obtained intermediate-resolution spectra of EPIC~204440603/2MASS~J16142312-2219338 from 5750 to 8800\AA, deriving a spectral type of M5.75 and H$\alpha$ equivalent width of -94.5\AA. Their measured Na~I and K~I gravity-sensitive equivalent widths as well as detection of Li~I absorption cements the classification of this object as a young low-mass star.

\subsection{EPIC 204830786 / 2MASS J16075796-2040087}

This Upper Scorpius member was first identified as a strong H$\alpha$ emission-line star by \citet{1964PASP...76..293T}. \citet{2009ApJ...703.1511K} obtained low-resolution spectra, which confirmed H$\alpha$ emission at an equivalent width of -357\AA\ and \ion{Ca}{2} triplet emission as well. Detection of further emission lines (\ion{N}{2}, \ion{S}{2}, \ion{Fe}{2}, \ion{Ni}{2}, \ion{O}{1}, and the Paschen series) suggested accretion-driven jets. These authors also associated EPIC~204830786/2MASS~J16075796-2040087 with a wide-separation companion some 21.5\arcsec\ (3120~AU) away. It has a significant infrared excess, as shown with 
IRAS \citep{1992A&A...262..106C} and later with {\em Spitzer} and WISE by \citet{2012ApJ...758...31L}.

Our HIRES spectrum suggests a spectral type of late G to early K but the spectrum is clearly heavily veiled; 
the lithium strength is $W_{Li}=0.20$ \AA. Strong H$\alpha$ emission is seen, as indicated in Table~\ref{tab:obstable}, with a blue-side absorption notch  in the profile. Strong and broad \ion{He}{1}, NaD,  \ion{Fe}{2}, \ion{Ca}{2}, 
\ion{O}{1} 8446, and perhaps other emission is present. Strong multi-component forbidden emission lines of 
[\ion{O}{1}] and [\ion{S}{2}]  are seen, along with single-component [\ion{N}{2}]  and many [\ion{Fe}{2}] lines. 

Our DSSI speckle observations do not identify any companions outward of 0.1\arcsec\ from this star, at a magnitude difference of $\Delta m\sim 4$ in the 692 and 880~nm bands. 

\subsection{EPIC 204906020 / 2MASS J16070211-2019387} \citet{2001AJ....121.1040P} detected EPIC~204906020 as a youthful member of the Upper Sco association via spectroscopic measurement of lithium absorption. This M5 star also has H$\alpha$ emission, although it was weak enough in some observations to lead to a weak-lined T Tauri star classification.  \citet{2007ApJ...662..413K} reported the object to be an M5/M5.5 wide binary with a 1.63\arcsec\ separation. \citet{2009ApJ...703.1511K} found that the primary is itself a binary, with 55~mas projected separation (8~AU at the distance of Upper Sco).

\citet{2006ApJ...651L..49C} reported infrared excesses at 8 and 16~$\mu$m, indicating a disk. \citet{2009ApJ...705.1646C} also detected an excess at 24~$\mu$m with {\em Spitzer}/MIPS. \citet{2012ApJ...745...23M} did not detect any cool dust around this system at millimeter wavelengths, at a 3-$\sigma$ upper limit of 3.7$\times 10^{-3}$~$M_{\rm Jup}$. 
However, \citet{2013A&A...558A..66M} used {\em Herschel}/PACS to detect 6.6$\times 10^{-6}$~$M_\odot$ worth of dust.
\citet{2014ApJ...787...42C} detected but did not resolve the disk with ALMA.

EPIC~204906020/2MASS~J16142312-2219338 is also known to have a circumstellar disk, as first reported by \citet{2012MNRAS.420.2497R} and confirmed by \citet{2012ApJ...758...31L}. The SED slope is -1.3, making it a class II disk \citep{2012MNRAS.420.2497R}.

Our speckle observations with DSSI rule out any companions beyond 0.1\arcsec\ at 4.6 magnitudes contrast in the 692 and 880~nm bands.

\subsection{EPIC 204908189 / 2MASS J16111330-2019029} 
This source appears in the literature only in the \citet{2012ApJ...758...31L} $WISE$ sample and the \citet{Baren16} ALMA study.
Our HIRES spectrum suggests a spectral type of M1 with lithium absorption present at strength $W_{Li}=0.15$ \AA. 
Strong H$\alpha$ emission is seen, as indicated in Table~\ref{tab:obstable}, along with \ion{He}{1}, weak but broad NaD, weak and narrow \ion{Fe}{2}, weak and narrow \ion{Ca}{2}, but moderately broad \ion{O}{1} 8446 emission.
Weak and narrow [\ion{O}{1}] is also present.

\subsection{EPIC 205008727 / 2MASS J16193570-1950426} 
There is no previous literature on this source. The HIRES spectrum is heavily veiled but appears to be a late K to M3 type, with lithium present at strength $W_{Li} = 0.55 A$. H$\alpha$ emission as reported in Table~\ref{tab:obstable} is strong and has multiple components. Additional emission includes \ion{He}{1}, NaD, \ion{Fe}{2}, \ion{Ca}{2} with the same profile shape as the H$\alpha$ and \ion{O}{1} 8446 is present.  Outflow lines of [\ion{O}{1}], [\ion{N}{2}], and [\ion{S}{2}] are also present, along with [\ion{Fe}{2}]. 

\subsection{EPIC 205061092 / 2MASS J16145178-1935402}
There is no previous literature on this source. Our HIRES data indicate a spectral type of M5-M6 with lithium at $W_{Li} = 0.59 A$. Strong H$\alpha$ emission, as indicated in Table 1, is present along with \ion{He}{1}, but no other emission lines.

\subsection{EPIC 205088645 / 2MASS J16111237-1927374}
\citet{2002AJ....124..404P} first identified this star as an M5 member of the USco association, based on lithium absorption and broad H$\alpha$ emission (-50\AA). \citet{2009A&A...504..981B} assigned it a membership probability of 99.9\% based on the USNO-B proper motion. \citet{2010A&A...517A..53M} found a slightly later spectral type of M6 based on low-resolution spectra, and similarly broad H$\alpha$ emission. The object displays an infrared excess confirmed by {\em WISE} to come from a full disk \citep{2012ApJ...758...31L}.

\subsection{EPIC 205156547 / 2MASS J16121242-1907191} 
There is no previous literature on this source. Our spectrum indicates a spectral type of M5-M6 with lithium at $W_{Li} = 0.56 A$. Moderate H$\alpha$ emission, as indicated in Table 1, is apparent with with an asymmetric extension on the blue side of the profile. There is also \ion{He}{1} but no other emission lines.

\section{Spectral Energy Distributions}

\begin{figure*}
\epsscale{1.10}
\plotone{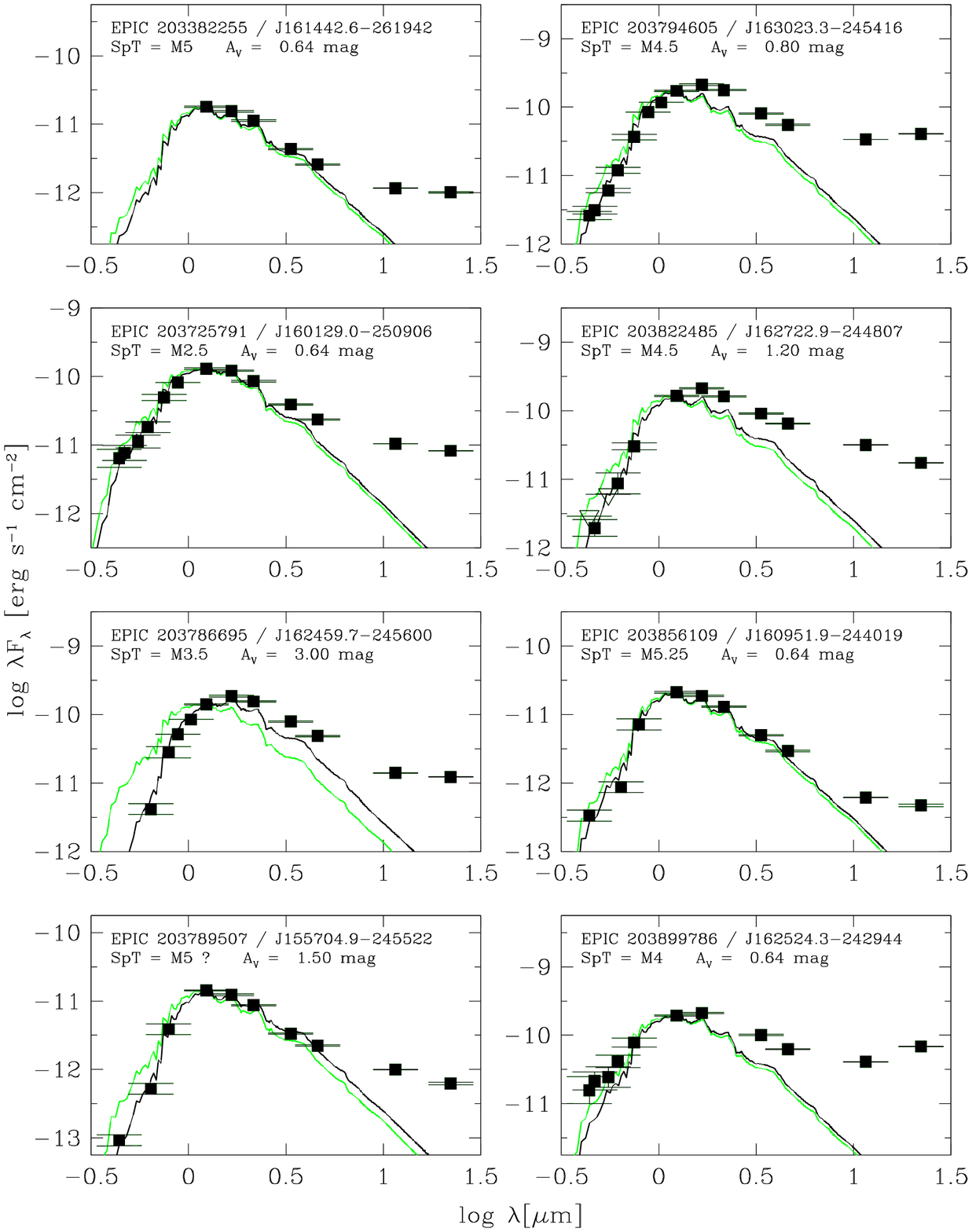}
\caption{Spectral Energy Distributions for all 29 sources studied in the paper.  
 Photometry has been assembled from 
 APASS (BV,gri), UKIDSS (ZY), 2MASS (JHK), and WISE (W1, W2, W3, W4).  The green line
 is a NextGen2 stellar atmosphere \citep{1999ApJ...512..377H} at log $g = 4.0$ and the temperature corresponding to
 the quoted spectral type.  The black line is the same photosphere reddened by the quoted $A_V$.
 A value of 0.64 mag -- the median $A_V$ we derive from assessment of extinction for
 several hundred members of the Upper Sco region -- has been adopted as a minimum, with
 higher values of $A_V$ used when needed in order to fit the optical and near-infrared SED.}
\label{seds}
\end{figure*}
 
\begin{figure*} 
 \addtocounter{figure}{-1}
\epsscale{1.10}
\plotone{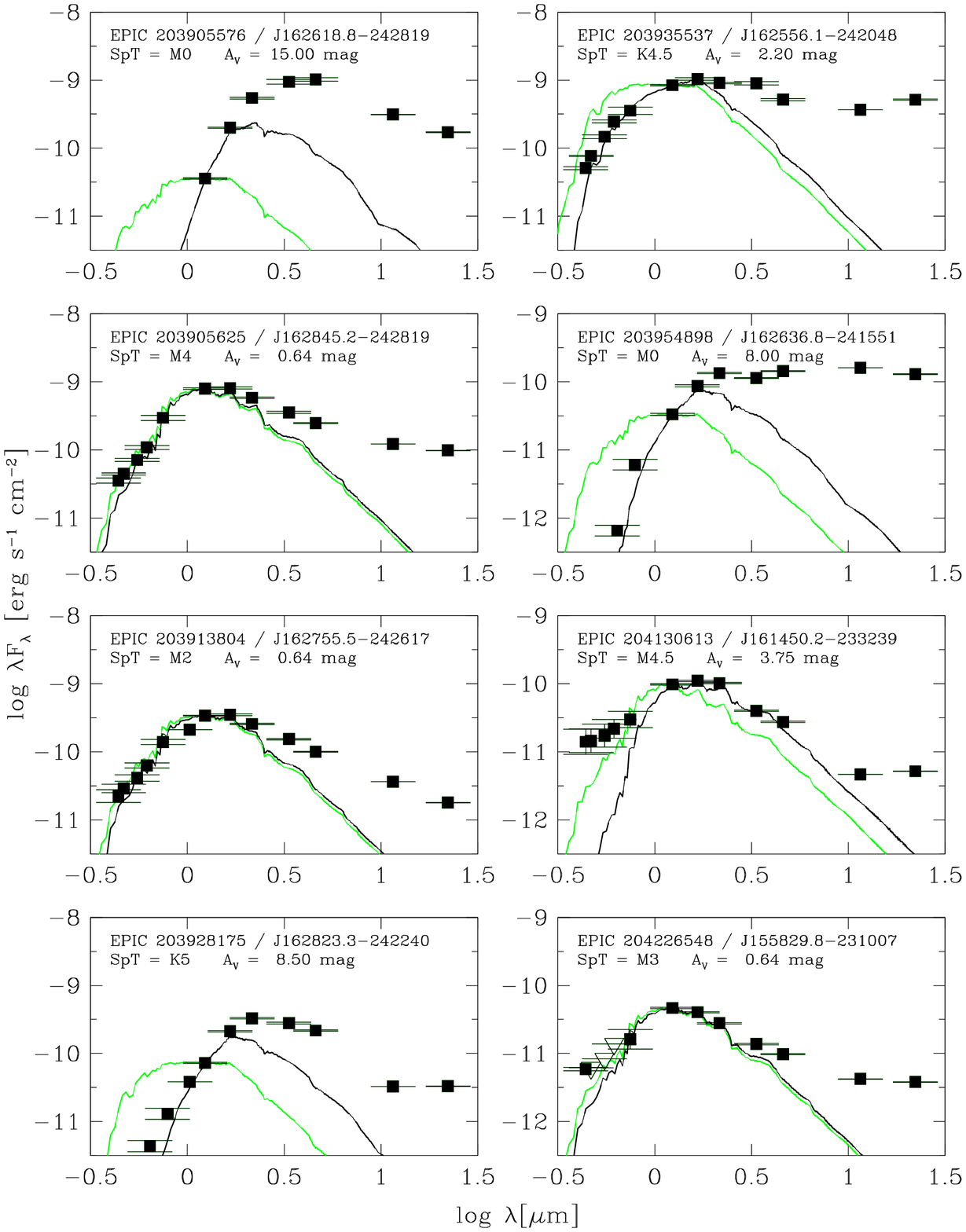}
\caption{Cont.}
\end{figure*}

\begin{figure*} 
\addtocounter{figure}{-1}
\epsscale{1.10}
\plotone{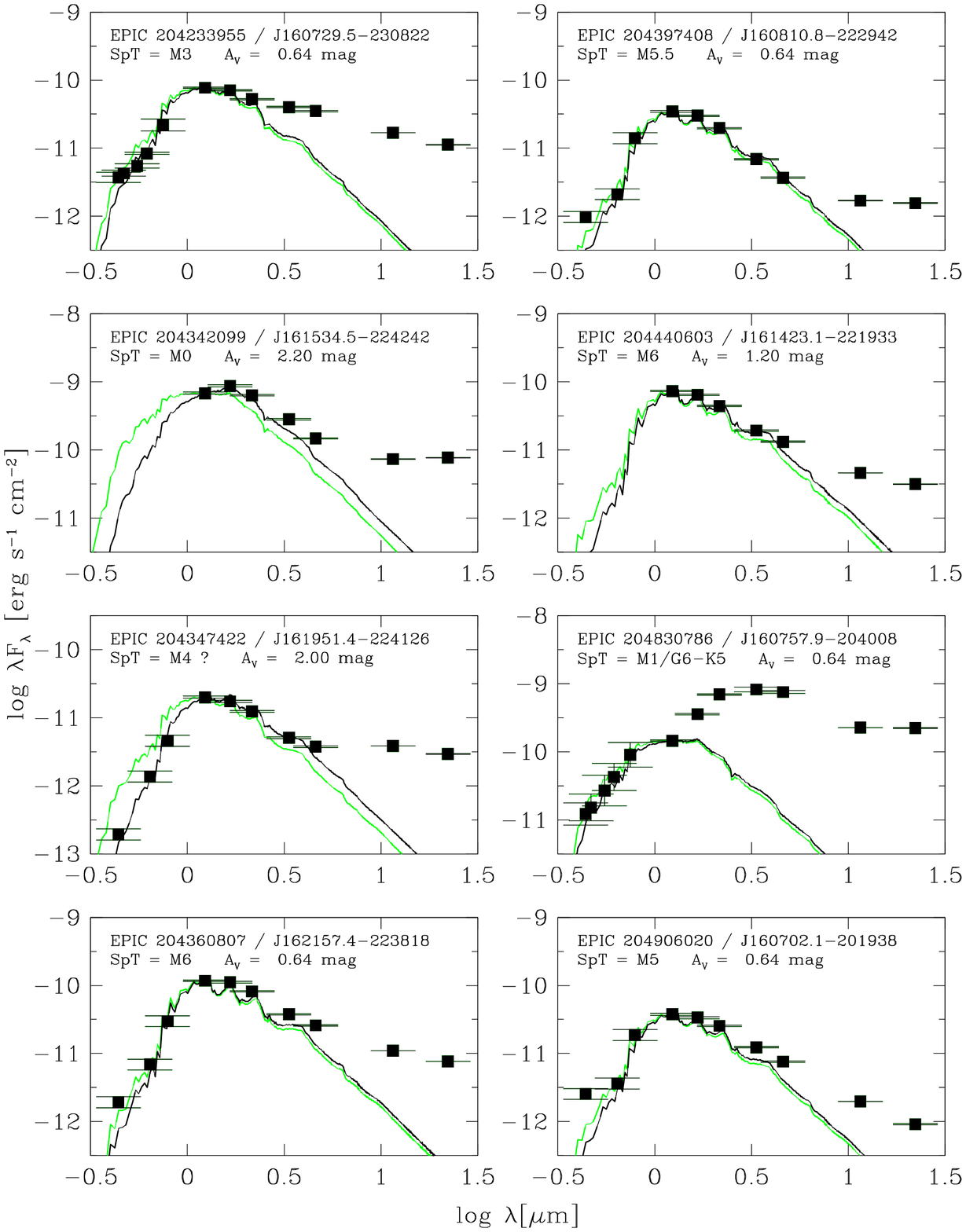}
\caption{Cont.}
\end{figure*}

\begin{figure*} 
\addtocounter{figure}{-1}
\epsscale{1.10}
\plotone{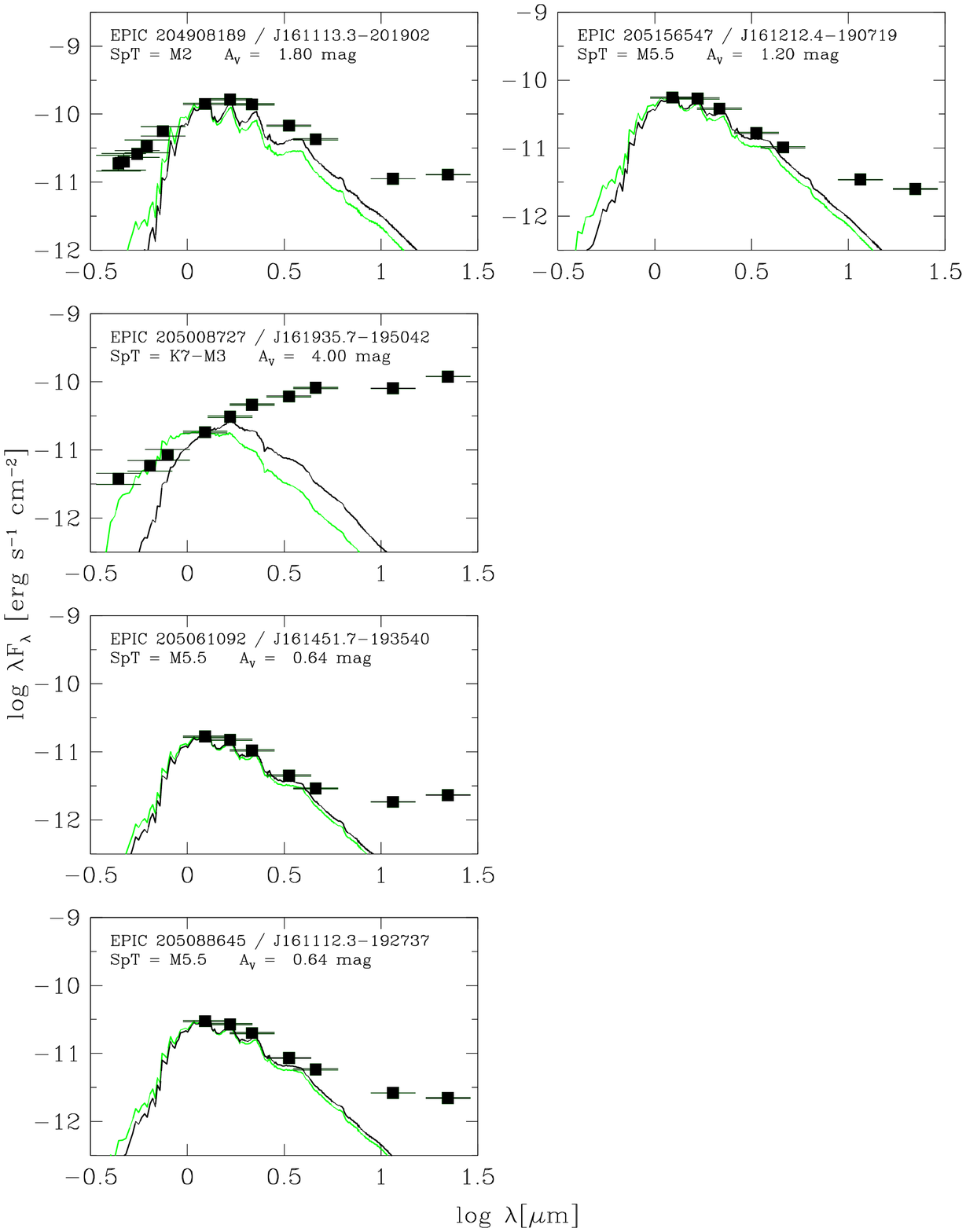}  
\textbf{}
\end{figure*}

\end{document}